\documentclass[]{emulateapj}
\usepackage[utf8]{inputenc}
\usepackage{amssymb,natbib,graphicx,color,ctable, amsmath,array}
\usepackage{hyperref}
\hypersetup{
    colorlinks=true,
    linkcolor=blue,
    citecolor=blue,
    filecolor=blue,
    urlcolor=blue
}
\usepackage[flushleft]{threeparttable}

\begin{document}
\title{\rm HI-to-H$_2$ Transitions in Dust-Free Interstellar Gas}

\author{Amiel Sternberg\altaffilmark{1, 2 ,3}$^{\star}$, Alon Gurman$^{1}$, \&
Shmuel Bialy\altaffilmark{4}}
\altaffiltext{1}{School of Physics and Astronomy, Tel Aviv University, Ramat Aviv 69978, Israel
 }
 \altaffiltext{2}{Center for Computational Astrophysics, 162 5th Ave., New York, NY, 10010
 }
 \altaffiltext{3}{Max-Planck-Institut f\"ur extraterrestrische Physik (MPE), Giessenbachstr., 85748 Garching, FRG
 }
\altaffiltext{4}{Center for Astrophysics $\vert$ Harvard \& Smithsonian, 60 Garden st., Cambridge, MA, 02138
 }
\email{$^\star$asternberg@flatironinstitute.org}
\slugcomment{ApJ. accepted}



\defcitealias{Bialy2016a}{BS16}
\defcitealias{Sternberg2014}{S14}
\defcitealias{Neufeld2020}{N20}
\defcitealias{McElroy2013}{M13}




\begin{abstract}
We present numerical computations and analysis of atomic to molecular (H{\small I}-to-H$_2$) transitions in cool ($\sim$100~K) low-metallicity dust-free (primordial) gas, in which molecule formation occurs via cosmic-ray driven negative ion chemistry, and removal is by a combination of far-UV photodissociation and cosmic-ray ionization and dissociation. For any gas temperature, the behavior depends on the ratio of the Lyman-Werner (LW) band FUV intensity to gas density, $I_{\rm LW}/n$, and the ratio of the cosmic-ray ionization rate to the gas density, $\zeta/n$. We present sets of H{\small I}-to-H$_2$ abundance profiles for a wide range of $\zeta/n$ and $I_{\rm LW}/n$, for dust-free gas. We determine the conditions for which H$_2$ absorption line self-shielding in optically thick clouds enables a transition from atomic to molecular form for ionization driven chemistry. We also examine the effects of cosmic-ray energy losses on the atomic and molecular density profiles and transition points. For a unit Galactic interstellar FUV field intensity ($I_{\rm LW}=1$) with LW flux $2.07\times 10^7$ photons cm$^{-2}$ s$^{-1}$, and a uniform cosmic-ray ionization rate $\zeta=10^{-16}$~s$^{-1}$, an H{\small I}-to-H$_2$ transition occurs at a total hydrogen gas column density of $4\times 10^{21}$~cm$^{-2}$, within $3\times 10^7$ yr, for a gas volume density of $n=10^6$~cm$^{-3}$ at 100~K. For these parameters, the dust-free limit obtains for a dust-to-gas ratio Z$^\prime_d \lesssim 10^{-5}$, which may be reached for overall metallicities $Z^\prime\lesssim 0.01$ relative to Galactic solar values.
\end{abstract}




\section{Introduction}
Atomic to molecular hydrogen (H{\small I}-to-H$_2$) transitions are of fundamental importance for the structure of the interstellar medium (ISM) in galaxies \citep[][hereafter \citetalias{Sternberg2014}]{Sternberg2014}. Numerous observational studies show that dense molecular gas is closely correlated with star-formation, from sub-pc to kpc scales, and across cosmic time from low to high redshifts \citep{Tacconi2020}. H$_2$ formation may promote cooling and cloud fragmentation, and/or it may be accelerated in gravitationally collapsing regions where the gas becomes dense, and optically thick to destructive radiation fields.

In star-forming galaxies, especially Milky Way like systems containing diffuse to dense interstellar gas clouds, it is well established that hydrogen molecules are generally formed on the surfaces of dust grains
\citep{Gould1963,Jura1974,LeBourlot2012,Girichidis2020}. The formation efficiencies (via chemisorption and/or physisorption) depend on the solid-state dust compositions, the gas and grain temperatures, and the overall gas-to-dust mass ratios \citep{Cuppen2010, Wakelam2017}.
For systems in which the heavy element and dust abundances are very low, or if the grain surface temperatures are too high, the grain induced formation rates vanish, and molecules are instead formed by gas phase processes, especially via the H$^-$ negative ion intermediary \citep{McDowell1961, deJong1972, Dalgarno1973, Glover2003, Larsson2012}.
For example, gas phase production of H$_2$ may dominate in protoplanetary disks where high temperatures inhibit gas-grain sticking \citep{Glassgold2004}, or in dust-poor outflow jets from protostars \citep{Tabone2020}. In metal rich disks and jets, gas phase H$_2$ formation enables synthesis of heavy molecules (CO, OH, etc.) via the usual neutral-neutral and ion-molecule reactions, as occurs in dusty clouds \citep{Herbst1973, Sternberg1995,Tielens2013,VanDishoeck2014,Bialy2015a,Oberg2021}.
Atomic to molecular conversion for dust-free conditions may also be important in low-metallicity damped Lyman-$\alpha$ absorbers (DLAs) \citep{Ranjan2018}
or perhaps in dense condensations within cool filaments in the circumgalactic medium (CGM) of galaxies and/or the cosmic web.

Gas phase H$_2$ formation was critically important in the dust-free early Universe where even minute amounts of H$_2$ enabled cooling to low ($\sim 100$~K) temperatures, reduction of the Jeans masses, and the collapse of the first objects \citep{Peebles1968, Tegmark1997, Lepp2002, Haiman2016}. 
Feedback in the form of photodissociating Lyman-Werner (LW) band radiation from the first stars plausibly regulated the early star-formation rates and the formation of seed black holes  \citep{Haiman1997, Visbal2014,  Latif2014, Wolcott-Green2021}.  
H{\small I}-to-H$_2$ conversion for Galactic conditions, and the structures of photodissociation regions (PDRs) in dense Milky Way molecular clouds are well understood
\citep{Federman1979,Tielens1985,Black1987, Rollig2006,McKee2010,Sternberg2014}.
But how do analogous H{\small I}-to-H$_2$ PDR transitions occur for early Universe conditions?

The astrophysics of H{\small I}-to-H$_2$ conversion in the dusty interstellar medium (ISM) was reviewed in \citetalias{Sternberg2014}. That paper also presented numerical computations and analytic theory for H{\small I}-to-H$_2$ transitions and the build up of photodissociated H{\small I} columns in dusty FUV-irradiated molecular cloud surfaces (PDRs). In \citet[][hereafter \citetalias{Bialy2016a}]{Bialy2016a} we presented an analytic procedure for generating depth-dependent H{\small I}/H$_2$ density profiles, again for dusty clouds, for a wide range of assumed gas-to-dust mass ratios. In all of these computations, dust plays a dual role, (a) in providing the molecular formation sites, and (b) as shielding agents via FUV absorption (and scattering). Both roles, especially shielding, enable the atomic to molecular conversions.

In this paper, we extend the methodology presented in \citetalias{Sternberg2014} and \citetalias{Bialy2016a} for a theoretical study of atomic-to-molecular transitions in dust-free environments. We consider simple idealized steady-state models consisting of FUV irradiated plane parallel isothermal constant density slabs, in which H$_2$ formation is via ionization driven gas phase processes, and attenuation of the photodissociation rate is by H$_2$ absorption-line self-shielding only. This as opposed to dusty clouds in which H$_2$ is formed on dust grains, and shielding includes dust absorption in addition to molecular self-shielding. We assume that the cloud ionization is provided by a flux of (low-energy) penetrating cosmic-ray particles. We do not consider irradiation by X-rays. We assume that the fractional ionizations are not affected by the presence of any heavy elements (``metals"), or by interactions with dust grains or large molecules such as polycyclic aromatic hydrocarbons (PAHs). Ionization drives the molecular hydrogen formation chemistry everywhere in the cloud, including the outer photodissociation regions (PDRs) and the inner optically thick cosmic-ray zones (CRZs).  Ionization also serves as an H$_2$ removal mechanism in the cosmic-ray zones. 
 We present analytic and numerical computations for the H{\small I} and H$_2$ density profiles for the dust-free conditions. 

In \S 2, we present our model ingredients. This includes a review of the metal-free hydrogen-helium chemical network, and a discussion of FUV photodissociation, H$_2$ self-shielding, and cosmic-ray propagation.
In \S 3 we present an analytic model for the expected atomic to molecular profiles, and dependence on the FUV field strength, gas density, and ionization rate. We also consider the time-scales. 
We develop expressions for the total H{\small I} column densities in the PDRs and CRZs, as functions of FUV radiation intensity, cosmic-ray ionization rate, gas density, and temperature. 
In \S 4 we present numerical computations for the atomic and molecular density profiles and transition points for dust free conditions.  These computations include a fiducial model and a paramter study for a range of FUV intensities, gas densities, and ionization rates. We compare to our analytic description. We summarize in \S 5.


\section{Model Ingredients}

\subsection{Chemistry}
{
\renewcommand\theequation{R\arabic{equation}}
\setcounter{equation}{0}

We consider a steady-state between formation of H$_2$ via cosmic-ray driven negative ion chemistry, and destruction by FUV photodissociation and cosmic-ray ionization and dissociation, in a dust-free environment. 

Our metals-free hydrogen and helium chemical networks are shown in Figure \ref{fig: chem_network}. The reactions and rate coefficients are listed in Tables \ref{table: chem}, 
and \ref{table: cr}. 
We solve for the steady state abundances of H, H$_2$, H$^-$, H$^+$, H$_2^+$, H$_3^+$, He, He$^+$, HeH$^+$, and electrons. No heavy elements are included.  The species abundances vary with cloud depth as the photodissociating FUV radiation is absorbed by the H$_2$, and as cosmic-ray energy losses reduce the ionization rates.

\begin{figure}
	
	\centering
	\includegraphics[width=1\columnwidth]{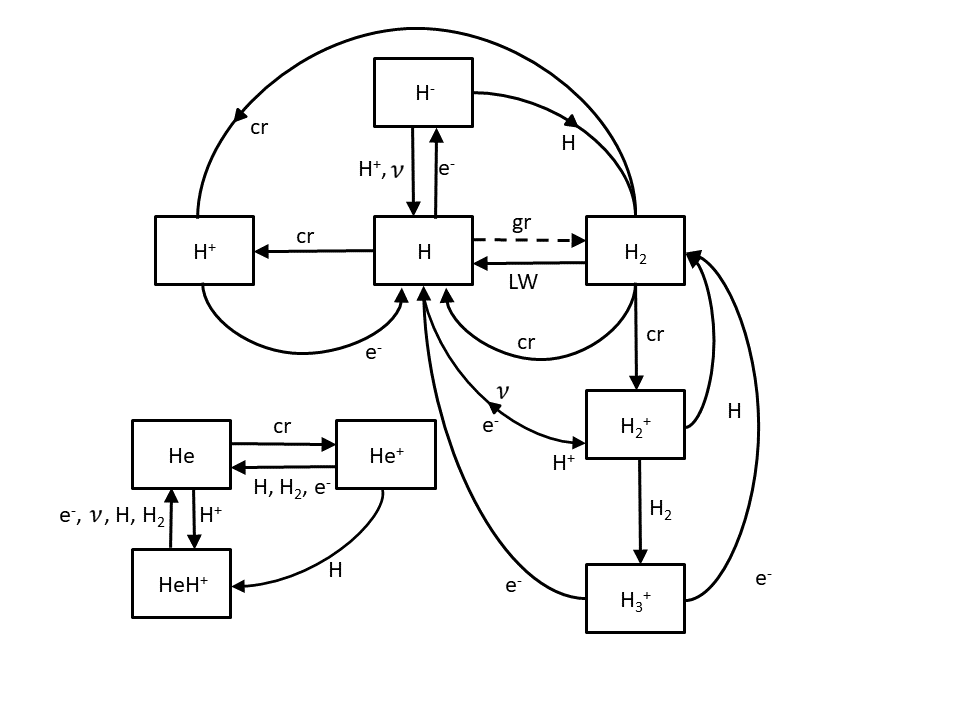} 
	\caption{
Schematic of the metals-free two-body reaction gas-phase hydrogen-helium network. The dashed arrow indicates grain-surface formation of H$_2$ that we are explicitly excluding.
		}
		\label{fig: chem_network}
\end{figure}

\begin{table*}
\centering 
\begin{threeparttable}
\caption{Rate coefficients}
\begin{tabular}{l l l l}
\hline \hline Reaction & Rate Coefficient $(\rm{cm}^3\;\rm{s}^{-1})$& Index & Reference\\ [0.5ex] \hline$\rm{H}\;+\;\rm{e}\rightarrow\rm{H}^-\;+\;\nu$ &$1.67\times 10^{-16}T_2^{0.64}e^{-0.092/T_2}$& R1 & \citetalias{McElroy2013}, \cite{Stancil1998}\\[0.5ex]
$\rm{H}^-\;+\;\rm{H}\rightarrow\rm{H}_2\;+\;\rm{e}$ &$6.63\times 10^{-9}T_2^{-0.39}e^{-0.394/T_2}$& R2 & \citetalias{McElroy2013}, \cite{Bruhns2010}\\[0.5ex]
$\rm{H}^-\;+\;\rm{H}^+\rightarrow\rm{H}\;+\;\rm{H}$ &$1.30\times 10^{-7}T_2^{-0.50}$& R3 & \citetalias{McElroy2013}, \cite{Harada2008}\\[0.5ex]
$\rm{H}^+\;+\;\rm{H}\rightarrow\rm{H}_2^+\;+\;\nu$ &$2.31\times 10^{-19}T_2^{1.50}$& R5 & \citetalias{Neufeld2020}, \cite{Ramaker1976}\\[0.5ex] 
$\rm{H}\;+\;\rm{H}_2^+\rightarrow\rm{H}_2\;+\;\rm{H}^+$ &$6.40\times 10^{-10}$& R6 & \citetalias{McElroy2013}, \cite{Karpas1979}\\[0.5ex]
$\rm{H}^+\;+\;\rm{e}\rightarrow\rm{H}\;+\;\nu$ &$7.98\times 10^{-12}T_2^{-0.75}$& R9 & \citetalias{McElroy2013}, \cite{Prasad1980b}\\[0.5ex]
$\rm{H}_2^+\;+\;\rm{H}_2\rightarrow\rm{H}_3 ^+\;+\;\rm{H}$ &$2.08\times 10^{-9}$& R10 & \citetalias{McElroy2013}, \cite{Theard1974}\\[0.5ex]$\rm{H}_2^+\;+\;\rm{e}\rightarrow\rm{H}\;+\;\rm{H}$ &$1.89\times 10^{-8}T_2^{-0.40}$& R11 & \citetalias{Neufeld2020}, \cite{Schneider1994}\\[0.5ex] 

$\rm{H}_3 ^+\;+\;\rm{e}\rightarrow\rm{H}\;+\;\rm{H}\;+\;\rm{H}$ &$7.72\times 10^{-8}T_2^{-0.52}$& R13 & \citetalias{McElroy2013}, \cite{McCall2004}\\[0.5ex]$
\rm{H}_3 ^+\;+\;\rm{e}\rightarrow\rm{H}_2\;+\;\rm{H}$ &$4.14\times 10^{-8}T_2^{-0.52}$& R14 & \citetalias{McElroy2013}, \cite{McCall2004}\\[0.5ex]
$\rm{H}\;+\;\rm{He}^+\rightarrow\rm{He}\;+\;\rm{H}^+$ &$9.12\times 10^{-16}T_2^{0.25}$& R21 & \citetalias{McElroy2013}, \cite{Stancil1998b}\\[0.5ex]
$\rm{H}_2\;+\;\rm{He}^+\rightarrow\rm{He}\;+\;\rm{H}_2^+$ &$7.20\times 10^{-15}$& R22 & \citetalias{McElroy2013}, \cite{Barlow1984}\\[0.5ex]
$\rm{H}_2\;+\;\rm{He}^+\rightarrow\rm{He}\;+\;\rm{H}\;+\;\rm{H}^+$ &$3.70\times 10^{-14}e^{-0.35/T_2}$& R23 & \citetalias{McElroy2013}, \cite{Barlow1984}\\[0.5ex]
$\rm{He}^+\;+\;\rm{e}\rightarrow\rm{He}\;+\;\nu$ &$9.28\times 10^{-12}T_2^{-0.50}$& R24 & \citetalias{McElroy2013}, \cite{Ercolano2006}\\[0.5ex]
$\rm{H}\;+\;\rm{He}^+\rightarrow \rm{HeH}^+\;+\;\nu$ & $1.44\times 10^{-16}$ & R25 & \citetalias{Neufeld2020}, based on \cite{Vranckx2013}\\[0.5ex] 
$\rm{H}^+\;+\;\rm{He}\rightarrow\rm{HeH}^+\;+\;\nu$ &$1.77\times 10^{-18}T_2^{-1.25}$& R26 & \citetalias{Neufeld2020}, \cite{Jurek1995}\\[0.5ex] 
$\rm{H}\;+\;\rm{HeH}^+\rightarrow\rm{He}\;+\;\rm{H}_2^+$ &$1.70\times 10^{-9}$& R27 & \citetalias{Neufeld2020}, \cite{Esposito2015}\\[0.5ex]
$\rm{H}_2\;+\;\rm{HeH}^+\rightarrow\rm{He}\;+\;\rm{H}_3 ^+$ &$1.50\times 10^{-9}$& R28 & \citetalias{McElroy2013}, \cite{Bohme1980}, \cite{herbst1975}\\[0.5ex]
$\rm{HeH}^+\;+\;\rm{e}\rightarrow\rm{He}\;+\;\rm{H}$ &$4.31\times 10^{-9}T_2^{-0.50}$& R29 & \citetalias{Neufeld2020}, \cite{Novotny2019}\\[0.5ex]

\hline

\end{tabular}
\begin{tablenotes}
\item 
Rate coefficients for the two-body reactions in our network, operating in the low-temperature ($T\lesssim 10^3$~K) regime. Here, $T_2=T/(100$~K). The rate coefficients are from the UMIST2012 compilation
\citep[M13;][]{McElroy2013}, or the more recent \citep[N20;][]{Neufeld2020} compilation mainly for the helium chemistry. We list the primary references as given by \citetalias{McElroy2013} and \citetalias{Neufeld2020}.
\end{tablenotes}
\label{table: chem} 
\end{threeparttable}
\end{table*}





\begin{table*}
\centering
\begin{threeparttable}
\caption{Cosmic ray rates
}
\centering 

\begin{tabular}{l l l}

\hline \hline 
Reaction & Relative Rate& Index \\ [0.5ex] 
\hline
$\rm{H}\;+\;\rm{cr}\rightarrow \rm{H}^+\;+\;\rm{e}$ &$0.46$ & R7
\\[0.5ex]$\rm{H}_2\;+\;\rm{cr}\rightarrow \rm{H_2^+}\;+\;\rm{e}$ &$0.96$ & R8
\\[0.5ex]$\rm{H}_2\;+\;\rm{cr}\rightarrow \rm{H}\;+\;\rm{H}$ &$0.7$ & R15
\\[0.5ex]
$\rm{H}_2\;+\;\rm{cr}\rightarrow \rm{H}\;+\;\rm{H}^+ \;+\;\rm{e}$ &$0.04$ & R16
\\[0.5ex]

$\rm{He}\;+\;\rm{cr}\rightarrow \rm{He}^+\;+\;\rm{e}$ &$0.5$ & R20
\\[0.5ex]
\hline

\end{tabular}
\begin{tablenotes}
\item
Cosmic ray reaction rates relative to the total H$_2$ cosmic-ray ionization rate ([R8]+[R16]).
\end{tablenotes}
\label{table: cr} 
\end{threeparttable}
\end{table*}

\subsubsection{Hydrogen Network}
Molecular hydrogen is formed mainly via the well-known two-step catalytic sequence, starting with (slow) radiative attachment, \begin{equation}
    {\rm H + e \rightarrow H^- + \nu}
\end{equation}
followed by (rapid) associative detachment 
\begin{equation}
    {\rm H^- + H \rightarrow H_2 +e} \ \ \ .
\end{equation}
In regions with large fractional ionizations and/or low gas densities this sequence may be moderated by mutual neutralization 
\begin{equation}
    {\rm H^- + H^+ \rightarrow H + H} \ \ \ ,
\end{equation}
and photodetachment
\begin{equation}
    {\rm H^- + \nu \rightarrow H + e} \ \ \ .
\end{equation}
Photodetachment is 
by radiation shortward of the 1.6~$\mu$m  ($E>$ 0.75 eV) infrared threshold. For the high gas density parameter space we consider in this paper, [R3] and [R4] are slow in comparison to [R2], and do not reduce the molecular formation efficiencies.
An additional H$_2$ formation route is radiative association
\begin{equation}
    {\rm H + H^+ \rightarrow H_2^+ + \nu}
\end{equation}
followed by charge transfer
\begin{equation}
    {\rm H_2^+ + H \rightarrow H_2 + H^+ \ \ \ .}
\end{equation}
This channel is generally a minor contributor.

The H$_2$ formation sequence, [R1] and [R2], requires electrons, and these are  
produced by an ionization source that we assume are non-thermal cosmic-ray particles, 
\begin{equation}
    {\rm H + cr \rightarrow H^+ + e}
\end{equation}
and
\begin{equation}
{\rm H_2 + cr \rightarrow H_2^+ + e} \ \ \ .
\end{equation}
Cosmic-ray ionization is also the main source of protons. The protons are removed by slow radiative recombination
\begin{equation}
    {\rm H^+ + e \rightarrow H + \nu} \ \ \ .
\end{equation}
The H$_2^+$ ions undergo rapid abstraction 
\begin{equation}
    {\rm H_2^+ + H_2 \rightarrow H_3^+ + H} \ \ \ .
\end{equation}
These reactions, together with charge transfer [R6] are the primary removal channels for the H$_2^+$ ions.
Dissociative recombination
\begin{equation}
    {\rm H_2^+ + e \rightarrow H + H} \ \ \ ,
\end{equation}
is a minor removal process unless the fractional ionizations become large. The H$_2^+$ may also be removed by far-UV photons
\begin{equation}
    {\rm H_2^+ + \nu} \rightarrow {\rm H + H^+} \ \ \ ,
\end{equation}
but this is a minor removal channel in our parameter space.

The saturated H$_3^+$ ions produced by [R10] recombine quickly with electrons
\begin{equation}
    {\rm H_3^+ + e \rightarrow H + H + H}
\end{equation}
and
\begin{equation}
    {\rm H_3^+ + e \rightarrow H_2 + H} \ \ \ 
\end{equation}
leading back to H or H$_2$.

Cosmic-ray ionization [R8], and cosmic-ray dissociation
\begin{equation}
    {\rm H_2 + cr \rightarrow H + H}
\end{equation}
are the main H$_2$ removal mechanisms at cloud depths that are fully shielded from photodissociating far-UV radiation. 
Dissociative ionization 
\begin{equation}
    {\rm H_2 + cr \rightarrow H + H^+ + e}
\end{equation}
is an additional (minor) removal channel. As we discuss further in our analysis below (\S 3), when the ionization, [R8], is followed by an abstraction, [R10], an additional H$_2$ molecule is removed for every H$_3^+$ ion that dissociatively recombines into three atoms, [R13].

A primary H$_2$ destruction process is photodissociation by 912-1108~\AA \ Lyman-Werner (LW) photons 
\begin{equation}
    {\rm H_2 + \nu_{\rm LW} \rightarrow H + H} \ \ \ ,
\end{equation}
in the $B ^1\Sigma_u - X ^1\Sigma^+_g$ and $C ^1\Pi_u - X ^1\Sigma^+_g$ absorption line systems. Photodissociation is generally the dominant H$_2$ removal mechanism unless self-shielding is very significant, as we discuss in \S 3 and \S 4. (We are neglecting dust absorption.)
We assume that any hydrogen ionizing ultraviolet (EUV) radiation is absorbed outside the predominantly neutral atomic/molecular medium we are studying. Furthermore, we do not consider penetrating X-rays here. Local X-ray ionization is similar in its effects to cosmic-ray ionization. For simplicity of following the ionization driven chemistry, we assume that the ionization is provided by just cosmic-ray impacts throughout.  

In sufficiently dense gas, H$_2$ is also produced by the three-body reaction
\begin{equation}
    {\rm H + H + H} \rightarrow {\rm H_2 + H} \ \ \ .
\end{equation}
However, this requires very dense gas $\gtrsim 10^8$~cm$^{-3}$, as may occur in collapsing clouds
\citep{Palla1983, Abel2002, Turk2012}
and we do not consider this regime here. 
H$_2$ is also removed by collisional dissociation
\begin{equation}
    {\rm H + H_2} \rightarrow {\rm H + H + H} \ \ \ ,
\end{equation}
however, this requires warm gas, with $T\gtrsim 1000$~K \citep{Martin1996}, which is also outside our modeling regime that we are restricting to cool/cold gas.

\subsubsection{Helium Network}
For completeness we include the metals-free helium chemistry. Helium atoms are ionized by cosmic-ray impact
\begin{equation}
    {\rm He + cr \rightarrow He^+ + e} \ \ \ ,
\end{equation}
and neutralization is by rapid charge transfer and dissociative charge transfer
\begin{align}
    {\rm He^+ + H} &\rightarrow {\rm He + H^+} \\
    {\rm He^+ + H_2}  & \rightarrow {\rm He + H_2^+}\\
    {\rm He^+ + H_2}  & \rightarrow {\rm He + H + H^+} \ \ \ .
\end{align}
Radiative recombination 
\begin{equation}
    {\rm He^+ + e \rightarrow He + \nu}
\end{equation}
is an additional minor removal channel. 

In cold predominantly neutral gas, the molecular ion HeH$^+$ is produced by the radiative associations
\begin{equation}
    {\rm H + He^+ \rightarrow HeH^+ + \nu}
\end{equation}
\begin{equation}
    {\rm H^+ + He  \rightarrow HeH^+ + \nu} \ \ \ ,
\end{equation}
and is removed by
\begin{align}
    {\rm HeH^+ + H} &\rightarrow {\rm He + H_2^+}, \\  
    {\rm HeH^+ + H_2}  &\rightarrow  {\rm He + H_3^+} \ \ \ .
\end{align}
In our models
dissociative recombination
\begin{equation}
    {\rm HeH^+ + e \rightarrow He + H} 
\end{equation}
is a very minor removal channel because the electron fractions remain low. Photodissociation
\begin{equation}
    {\rm HeH^+ + \nu} \rightarrow {\rm He + H^+}
\end{equation}
is also a minor removal channel in our parameter space.

Reactions [R25] and [R26] \footnote{Reaction [R25] dominates the formation of HeH$^+$ in the partially ionized interface zones in planetary nebulae where removal by [R29] is important in addition to [R27]. Interstellar HeH$^+$ was recently observed for the first time in the planetary nebula NGC7027 \citep{Gusten2019, Neufeld2020}. In such nebulae, associative ionization ${\rm H + He(2^3S) \rightarrow HeH^+ + e}$ involving the metastable helium produced via recombination, is also an important source of HeH$^+$ \citep{Roberge1982, Neufeld2020}.}
followed by [R27] and [R6], as well as [R5] followed by [R6], were primary sources of H$_2$ following cosmic recombination and prior to the introduction of the first reionization sources. At redshifts $z\gtrsim 100$, and at the prevailing mean gas densities, any H$^-$ produced via relic electrons ions were rapidly photodetached by thermal background photons  \citep{Lepp2002, Galli2013}. 
In the computations we consider here, we assume efficient formation of H$^-$ via free electrons produced by local sources of ionizing cosmic-rays, and the H$_2$ is formed mainly by [R1] and [R2]. The main parameters controlling the chemistry are the Lyman-Werner radiation intensity and associated photodissociation rate, the cosmic-ray ionization rate, the gas density, and the cloud temperature. We assume that photodetachment is by the same radiation source that also photodissociates the H$_2$. For our parameter space, and relatively dense gas conditions, photodetachment is therefore negligible, and our results are insensitive to the IR to UV spectral shapes as we discuss below.

\subsection{FUV radiation}
\renewcommand\theequation{\arabic{equation}}
\setcounter{equation}{0}

In our study we compute depth dependent H{\small I} and H$_2$ density profiles for dust-free clouds exposed to fluxes of far-UV photons and cosmic-ray particles. We 
determine conditions for which an H{\small I}-to-H$_2$ transition occurs assuming steady-state conditions.

\subsubsection{Radiation Spectra}

We parameterize the far-UV radiation flux, in particular the dissociating Lyman-Werner band flux, relative to the standard Draine (1978) Galactic FUV interstellar field strength \citep{Parravano2003,Bialy2020b}.

For the Draine far-UV spectrum, the flux in the (narrow) 912-1108~\AA \ LW band is 
$F_{\rm LW}=2.07\times 10^7 I_{\rm LW}$~photons cm$^{-2}$ s$^{-1}$. $I_{\rm LW}$ is the radiation field strength\footnote{In \citetalias{Sternberg2014} we refer to the intensity parameter as $I_{\rm UV}$, and normalize to the entire 6-13.6 eV FUV band. Here we use the subscript ``LW" to emphasize normalization relative to just the Lyman-Werner band.
}
relative to the free-space unattenuated Draine field in the LW band, for which $I_{\rm LW}=1$.  The unattenuated H$_2$ photodissociation rate due to absorptions in all of the LW lines is $D_0=\sigma_d{\bar F}_{\nu,{\rm LW}}=5.8\times 10^{-11}$ s$^{-1}I_{\rm LW}$. Here 
${\bar F}_{\nu,{\rm LW}}$ is the mean flux-density (photons cm$^{-2}$ s$^{-1}$ Hz$^{-1}$) in the LW band, and 
$\sigma_d=2.36\times 10^{-3}$ cm$^2$ Hz is the total H$_2$-line photodissociation cross section, summed over all the LW absorption lines (see \citetalias{Sternberg2014}). For the Draine spectrum, the H$^-$ photodetachment rate is $D_-=5.6\times 10^{-9} I_{\rm LW}$~s$^{-1}$, and $D_-/D_0=96.6$.
We computed the photodetachment rate using the \cite{Miyake2010} cross section (see their Fig.~1) and integrating from the infrared threshold (1.6 $\mu$m) to the Lyman limit  (912~\AA). 

More generally, the ratio $D_-/D_0$ depends on the shape of the radiation spectrum from the near-IR to the UV \citep{Wolcott-Green2017}.  In Figure \ref{fig: uv_fields} we plot four radiation spectra, for pure Draine, combination Draine+MathisIR \citep{Mathis1983}, and $10^4$ and $10^5$~K blackbody fields\footnote{\cite{Mathis1983} represented their inferred IR Galactic stellar spectrum as a sum of diluted blackbodies at 3000, 4000, and 7500K, representing the late-type stars. The \cite{Draine1978} field is due primarily to the early OB type stars.}. We normalize the spectra in Figure \ref{fig: uv_fields} such that they all have the same flux of $2.07\times 10^7$~photons cm$^{-2}$ s$^{-1}$
in the LW band, appropriate for $I_{\rm LW}=1$. In Table \ref{table: D_} we list $J_{21}$ for these fields, defined as the flux densities at the Lyman limit in units of $10^{-21}$ erg cm$^2$ s$^{-1}$ Hz$^{-1}$ sr$^{-1}$.

\begin{figure*}
	\centering
	\includegraphics[width=2\columnwidth]{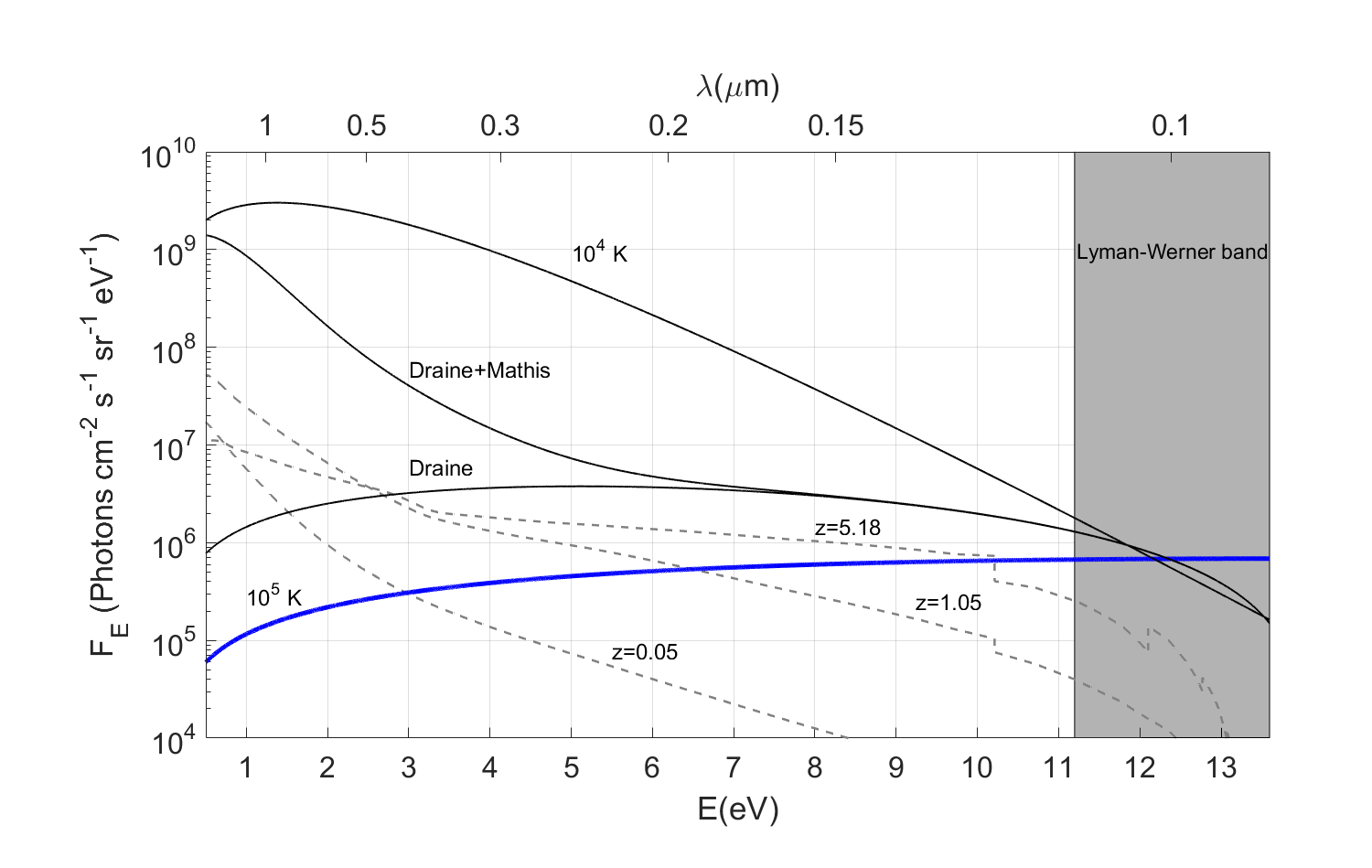} 
	\caption{
Solid curves show the (normalized) stellar-based radiation fields, Draine, Draine+Mathis, and 10$^4$ and 10$^5$~K blackbody spectra, all normalized such that $I_{\rm LW}=1$ in the Lyman-Werner band (shaded region). The dashed curves are the absolute (unnormalized) \cite{Haardt2012} metagalactic background radiation fields from low- to high redshift.
		}
		\label{fig: uv_fields}
\end{figure*}

\begin{table*}
\centering 

\begin{threeparttable}
\caption{Photo-Rates for stellar-based radiation fields}
\begin{tabular}{l l l l l}
\hline\hline 
Stellar Based \; \; \; \; \; \; &  $J_{21}$ $^{(a)}$  \; \; & $D_-(\rm{s}^{-1})$ $^{(b)}$ \; \;  & $D_-/D_0$ $^{(c)}$ \; \;&  $(I_{LW}/n_6)_{\rm crit}$ $^{(d)}$ 
  \\ [0.5ex]
   \hline 
Draine & 13.5& $5.58\times 10^{-9}$ & $96.6$  & $7.28\times10^5$\\

Draine+Mathis & 13.5&  $2.73\times 10^{-7}$ & $4.66\times 10^3$  & $6.88\times10^3$\\

$10^4\;\rm{K}$ & 14.6&  $1.77\times 10^{-7}$ & $3.05\times 10^3$ & $1.14\times10^4$\\

$10^5\;\rm{K}$  & 61.8&  $2.77\times 10^{-9}$ & $47.8$ & $1.56\times10^6$\\
     [0.5ex]
   \hline 

\end{tabular}
\begin{tablenotes}
      \footnotesize
\item $^{(a)}$ The radiation flux densities at the Lyman limit, in units of $10^{-21}$~ergs~s$^{-1}$~cm$^{-2}$~Hz$^{-1}$~sr$^{-1}$, assuming each field is normalized to $I_{\rm LW}=1$. 
\item $^{(b)}$ The H$^-$  photodetachment rates for the normalized fields. 
\item $^{(c)}$ The ratio $D_-/D_0$ of the H$^-$  photodetachment rate and H$_2$ photodissociation rate.
\item $^{(d)}$ The critical ratios $(I_{\rm LW}/n_6)_{\rm crit}$ above
which photodetachment reduces the H$^-$ abundance (see text). 
\end{tablenotes}
\label{table: D_} 
\end{threeparttable}
\end{table*}

\begin{table*}
\centering
\begin{threeparttable}
\caption{Photo-Rates for metagalactic fields}
\centering 
\begin{tabular}{l l l l l l l}
\hline\hline 
Metagalactic \;  \; & $J_{21}$ $^{(a)}$  \;  \; &$I_{\rm{LW}}$ $^{(b)}$ \;  \; & $D_-(\rm{s}^{-1})$ $^{(c)}$  \;  \; & $D_-/D_0$ $^{(d)}$  \;  \; &  $\big{(}n_6\big{)}_{\rm{crit}}$ $^{(e)}$  \;  \; & $(n/n_b)_{\rm{crit}}$ $^{(f)}$ \;  \;
  \\ [0.5ex]
   \hline 
$z=0.05$ & $8.24\times 10^{-3}$ &$7.00\times10^{-4}$& $1.70\times 10^{-9}$ & $4.19\times10^4$ &  $3.9\times10^{-7}$ & $1.36\times 10^6$\\

$z=1.05$ & 0.141 &$2.67 \times 10^{-2}$& $9.54\times 10^{-9}$ & $8.37\times10^3$ &  $2.72\times10^{-6}$ & $1.27\times 10^6$\\

$z=5.18$ &0.188 &$1.26 \times 10^{-1}$&  $6.16\times 10^{-9}$ & $8.44 \times 10^2$ &  $1.61\times10^{-6}$ & $2.75\times 10^4$\\
     [0.5ex]
   \hline 

\end{tabular}
\begin{tablenotes}
      \footnotesize
\item $^{(a)}$ Flux density $J_{21}$ at the Lyman limit. 
\item $^{(b)}$ LW intensity. 
\item $^{(c)}$ H$^-$ photodetachment rate. 
\item $^{(d)}$ The ratio $D_-/D_0$ of the H$^-$ photodetachment rate and H$_2$ photodissociation rate. 
\item $^{(e)}$ The critical density $n_{6,\rm crit}$ below which photodetachment reduces the H$^-$ abundance (see text).
\item $^{(f)}$ The critical photodetachment density relative to the mean cosmic baryon density $n_b=2.48\times 10^{-7}(1+z)^3$~cm$^{-3}$.
\end{tablenotes}
    \label{table: D_ metagalactic} 
\end{threeparttable}
\end{table*}

Because the LW band is narrow, we make the simplifying approximation that for equal LW fluxes the unattenuated H$_2$ dissociation rate is independent of the spectral shape and equal to $D_0=5.8\times 10^{-11}I_{\rm LW}$~s$^{-1}$. However, the H$^-$ photodetachement rate is very sensitive to the spectral shape. In Table \ref{table: D_} we list $D_-$ and $D_-/D_0$ for the four radiation fields. In Figure \ref{fig: D_ bb} we plot these quantities as functions of the radiation temperature, $T_{\rm rad}$, for pure blackbody fields, all normalized such that $I_{rm LW}=1$ at any $T{\rm rad}$. Figure \ref{fig: D_ bb} shows that $D_-/D_0$ grows rapidly for $T_{\rm rad}$ below $\sim 2\times 10^4$~K. In the computations in this paper we assume 10$^5$~K spectra, appropriate for massive pop~III stars, for which 
$D_-=2.7\times 10^{-9} I_{\rm LW}$ s$^{-1}$, and $D_-/D_0=46.5$. But even for this large ratio photodetachment does not play a role in our model parameter space, as we discuss further below.

For reference, in Figure \ref{fig: uv_fields} we also plot the expected cosmological metagalactic radiation spectra (unnormalized) computed by \cite{Haardt2012}, at redshifts $z$=0.05, 1.05, and 5.18. In Table \ref{table: D_ metagalactic} we list $J_{21}$, $I_{\rm LW}$, $D_-$, and $D_-/D_0$, for these three cosmological radiation fields.

\begin{figure}
	\centering
	\includegraphics[width=1\columnwidth]{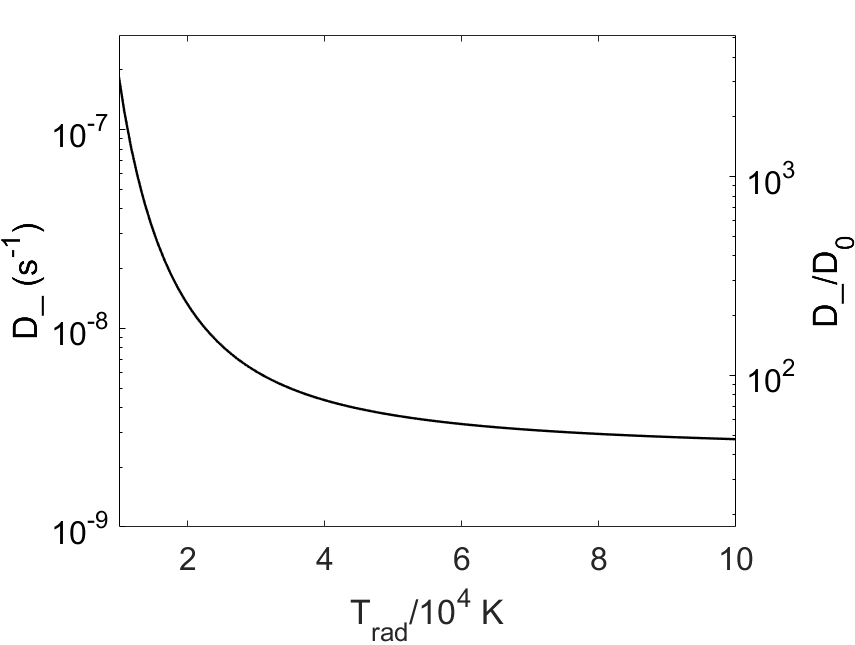} 
	\caption{
The H$^-$ photodetachment rate (left axis) and ratio of the H$^-$ photodetachment and H$_2$ photodissociation rates (right axis) as functions of radiation temperature for black body spectra, normalized to $I_{\rm LW}=1$ for any $T_{\rm rad}$.
		}
		\label{fig: D_ bb}
\end{figure}

\subsubsection{Shielding}

We are considering dust-free gas for which any attenuation of the H$_2$ photodissociation rate is due entirely to self-shielding optical depth in the H$_2$ absorption lines. In the dust-free regime the photodissociation rate at any depth into a semi-infinite plane-parallel slab is 
\begin{equation}
    D(N) = \frac{1}{2}D_0 f_{\rm shield}(N_{\rm H_2}) \ \ \ .
\end{equation}
Here, $N=N_{\rm HI}+2N_{\rm H_2}$  is the total column density of hydrogen nuclei, atomic (H{\small I}) plus molecular (H$_2$), from the surface to a depth, $z$ (cm), in the cloud. The factor of 1/2 accounts for reduction of the rate relative to free-space by the cloud itself, with radiation incident from one hemisphere only, and opaque to any radiation from the other side. This is the ``semi-infinite" limit. In this paper we assume beamed fields with incident rays all normal to the cloud surface\footnote{See \citetalias{Sternberg2014} for a discussion of beamed versus isotropic irradiation.}.

The H$_2$ self-shielding function is defined by
\begin{equation}
\label{eq: shield_def}
    f_{\rm shield}(N_{\rm H_2})\equiv \frac{1}{\sigma_d}\frac{dW_d(N_{\rm H_2})}{dN_{\rm H_2}} \ \ \ .
\end{equation}
In this expression, $W_d(N_{\rm H_2})$ is the bandwidth (Hz) of LW radiation absorbed in photodissociations, through all of the absorption lines, up to a molecular column $N_{\rm H_2}$ in a dust-free cloud, and $\sigma_d$ is the total H$_2$-line photodissociation cross section. For cool ($\lesssim 500$~K) gas the self-shielding function is well-approximated by  \citep[\citetalias{Sternberg2014}]{Draine1996}

\begin{align}
\label{eq: shield}
    f_{\rm shield} = \frac{0.965}{(1+ x/b_5)^2}
    &+ \frac{0.035}{(1+x)^{0.5}}  \\ \nonumber
    &\times \ {\rm exp}[-8.5\times 10^{-4}(1+x)^{0.5}]
\end{align}
where $x\equiv N_{\rm H_2}/(5\times 10^{14}\ {\rm cm}^{-2})$ and $b_5\equiv b/(10^5 {\rm cm \ s^{-1}})$ is the normalized absorption-line Doppler parameter. Alternate expressions have also been presented, 
especially for warmer material \citep[e.g.,][]{Wolcott-Green2011}. Here we adopt Eq.~(\ref{eq: shield}).

For very small H$_2$ column density, $f_{\rm shield}=1$. The onset of self-shielding occurs for $N_{\rm H_2}\sim 10^{14}$ cm$^{-2}$, depending on $b$ (we set $b_5=2$). As discussed by \citet[][see their Fig.~is 5]{Bialy2016a}, the atomic-to-molecular transition points are insensitive to the Doppler widths.
For $N_{\rm H_2} \gtrsim 10^{17}$ cm$^{-2}$ the absorptions occur out of the line damping wings and $f_{\rm shield}$ becomes small,
decreasing to $\sim 10^{-5}$ for $N_{\rm H_2}\sim 10^{21}$ cm$^{-2}$
and the dissociation rate is very significantly reduced.
For $N_{\rm H_2}\gtrsim 10^{22}$ cm$^{-2}$ the absorption lines overlap and the LW band is completely blocked \citep{Draine1996, Sternberg2014}. In this limit, $f_{\rm shield} \rightarrow 0$ and the photodissociation rate becomes vanishingly small. Self-shielding including line overlap\footnote{Line overlap occurs when the dust abundance and associated continuum opacity are small enough such that LW photons are absorbed in the outer Lorentzian damping wings of adjacent lines, prior to any dust absorption. This generally requires a LW dust-grain absorption cross section per hydrogen nucleus $\sigma_g\lesssim 10^{-22}$~cm$^{-2}$, for which the associated dust opacity $2\sigma_gN_{\rm H_2}\lesssim 1$.} enables an H{\small I}-to-H$_2$ transition even in the absence of dust shielding as we discuss below. 

The dissociation bandwidth in Eq.~(\ref{eq: shield_def}) is given by the fit formula\footnote{In \cite{Bialy2016a} our expression for the H$_2$ dissociation bandwidth also includes a dependence on the assumed dust absorption cross section. Here we are assuming $\sigma_g=0$.} \citep{Bialy2016a},

\begin{equation}
\label{eq: Wfit}
W_d(N_{\rm H_2}) \ = \ a_1 \ln \Big[ \frac{a_2 + y}{1 + y/a_3} \Big] \ \Big( \frac{1+y/a_3}{1+y/a_4} \Big)^{0.4} \ ,
\end{equation}
where 
\begin{align*}
y \ \ &\equiv \ \frac{N_{\rm H_2}}{10^{14} \ {\rm cm^{-2}}} \nonumber\\ 
a_1 \  &= \  3.6 \times 10^{11} \ {\rm Hz} \nonumber\\
a_2 \  &= \  0.62 \nonumber\\
a_3 \  &= \  2.6 \times 10^{3}  \nonumber\\
a_4 \  &= \  1.4 \times 10^7  \ \ \ .
\end{align*} 
Equation~(\ref{eq: Wfit}) is valid for $y\ge 1$.
For complete line absorption of the dissociating radiation in the LW band, $W_{\rm d,tot}=8.8\times 10^{13}$~Hz, and the total dissociation flux absorbed in an optically thick cloud is ${\bar F}_\nu W_{\rm d,tot}=2.2\times 10^6I_{\rm LW}$ photons cm$^{-2}$ s$^{-1}$. This is 10\% of the total LW flux, and is reradiated as dissociation continuum. The remaining energy is reradiated as Lyman-Werner band emission lines \citep{Sternberg1989a} terminating in mainly excited vibrational levels of the ground electronic state.

\subsection{Cosmic Rays}

Low level ionization in the shielded ISM of galaxies is generally maintained by low-energy cosmic rays.
\citep{Spitzer1968a, Dalgarno2006}. 
In localized environments photoionization by penetrating X-rays may also contribute in addition to the cosmic-ray impacts. Within molecular clouds 
cosmic-ray ionization drives a rich ion-molecule chemistry, and controls the dynamical coupling of magnetic fields to the gas.  In the Milky Way, the characteristic H$_2$ cosmic-ray ionization rate is $\sim 10^{-16}$ s$^{-1}$, although the rates may vary with location in the Galaxy and with cloud depth. \citep{Padovani2009,Neufeld2017b}. The cosmic-ray fluxes and ionization rates may be substantially larger in galaxies with elevated star-formation activity and FUV fields \citep{Mashian2013,Bialy2015a,Bisbas2017,Bialy2020b}. 

 Our parameter $\zeta$ is the total cosmic-ray H$_2$ ionization rate (s$^{-1}$) due to non-dissociative ([R8]) plus dissociative ([R15]) ionizations. We define the  normalized rate $\zeta_{-16}\equiv \zeta/(10^{-16}\ {\rm s}^{-1})$. The relative rates of all of the cosmic-ray ionization and dissociation processes in our network are listed in Table \ref{table: cr}. An important parameter for the chemistry is the number of H$_2$ dissociations per CR ionization, including the effects of the secondary electrons for both processes \citep{Cravens1978, Li2003, Goldsmith2005}.  Following \cite{Padovani2018} we assume 0.7  dissociations per ionization. This is the asymptotic ratio for large penetration depths.

 The penetration and transport of cosmic-rays, and the possible attenuation of the associated ionization rates with cloud depth, depend on a wide range of (uncertain) properties including the spectrum of the low energy particles, the distributions and orientations of the magnetic fields, the particle gyroradii, free-streaming ionization energy losses, particle production, and the characteristics of MHD turbulent diffusion
 \citep{Ivlev2018,Padovani2020}. 
 
 We therefore construct two families of models for (a)  constant ionization rates through the clouds, and (b) attenuation of the ionization rates assuming a power-law form for the energy losses. 
For model family (b), we adopt the \citet{Padovani2018a} results for the particle energy losses and transport for the ``Voyager-1 based" cosmic-ray input spectrum \citep{Cummings2016}.
  The \citet{Padovani2018a} CR ionization rates, for low- and high-ionization cosmic-ray spectra (their models ${\cal L}$ and ${\cal H}$), 
  may be expressed as simple power-laws
  \begin{equation}
  \label{eq: Pfit}
 \zeta=\zeta_0\times \biggl(\frac{N_{\rm eff}}{10^{19}\ {\rm cm}^{-2}}\biggr)^{-q} \ \ \ ,
 \end{equation}
 where $N_{\rm eff}$ is the effective cosmic-ray absorbing gas column. The effective column depends on the orientation angle $\theta$ of the magnetic field along which the cosmic-ray protons propagate relative to the cloud normal, so that $N_{\rm eff}=N/\mu$, where $\mu\equiv cos\theta$. 
 Equation (\ref{eq: Pfit}) is valid for $N_{\rm eff}$
 between 10$^{19}$ and 10$^{24}$ cm$^{-2}$, the relevant range for our considerations. 
 We find that $q=0.280$ and $q=0.385$ for the Padovani models ${\cal L}$ and ${\cal H}$ respectively, with an accuracy of 37\% and 6\%. For $N_{\rm eff}<10^{19}$  cm$^{-2}$ we assume $\zeta=\zeta_0$.

}
\section{Analysis}

In \S 4 we present our numerical solutions to the chemical rate equations and resulting H{\small I}/H$_2$ density profiles, including computations of the atomic to molecular transition points, for the dust-free conditions we are considering. Before presenting our numerical calculations we first provide an analytic description as a guide.

\subsection{H{\small I}/H$_2$ Balance}
In a steady state, we write the H$_2$ formation-destruction equation as
\begin{equation}
\label{eq: form-des}
    R_-nx_{\rm HI} = \big\lbrace\frac{1}{2}D_0f_{\rm shield}(N_{\rm H_2}) + [0.96 f_d(1+f_3) + 0.7 + 0.04]\zeta\big\rbrace x_{\rm H_2}
\end{equation}
In this expression, $x_{\rm HI}\equiv n_{\rm HI}/n$ and $x_{\rm H_2}\equiv n_{\rm H_2}/n$ are the depth dependent atomic (H{\small I}) and molecular (H$_2$) abundance fractions, and $n=n_{\rm HI}+2n_{\rm H_2}$ is the total hydrogen volume density. We are assuming that the additional constituents all have very low abundances compared to H{\small I} or H$_2$. On the left-hand side, $R_-$ is the effective rate coefficient for H$_2$ formation via the H$^-$ sequence, [R1] and [R2] \citep{Bialy2019}. The first term on the right-hand side is the depth dependent photodissociation rate in the photodissociation region (PDR). It vanishes in the inner cosmic-ray zone (CRZ)
where cosmic-ray ionization and dissociation become the main H$_2$ removal mechanisms. These are included in the second term in brackets multiplied by $\zeta$. The numerical factors are the relative rates for [R8], [R14], and [R15] respectively (see Table \ref{table: cr}).

In Eq.(\ref{eq: form-des}), the factor 0.96$\zeta$ for cosmic-ray ionization is multiplied by the branching ratio
\begin{equation}
\label{eq: fd}
   f_d = \frac{k_{10}x_{\rm H_2}}{k_{10}x_{\rm H_2}+k_6x_{\rm HI}} \ \ \ .
\end{equation}
This is the probability that the production of H$_2^+$ via ionization is followed by the abstraction reaction [R10] to H$_3^+$ 
rather than charge transfer [R6] back to H$_2$ \citep[see Eq 15 in][]{Bialy2015a}. Dissociative recombination of H$_2^+$ [R11] may be ignored because the fractional ionizations are small.
The additional factor
\begin{equation}
    1 + f_3 = 1 + \frac{k_{13}}{k_{13}+k_{14}} \ \ \ ,
\end{equation}
where $f_3=0.65$ 
is the branching ratio for dissociative recombination fragmentation of H$_3^+$ into three atoms, [R13], rather than back to H$_2$, [R14] (see Fig. \ref{fig: chem_network}). The product, $(1+f_3)f_d$ is then the net number of H$_2$ molecules removed per ionization event [R8].
This number is depth dependent because $f_d$ depends on the atomic and molecular factions, $x_{\rm HI}$ and $x_{\rm H_2}$.  Thus, for $x_{\rm HI}\approx 1$
, e.g. in photodissociated gas,
$f_d=0$ and cosmic-ray ionization [R8] is ineffective in further reducing the small molecular fraction.  For $x_{\rm H_2}\approx 0.5$,
the branching ratio $f_d=1$, and 1.65 hydrogen molecules are removed per cosmic-ray ionization event. 
Cosmic-ray dissociation [R15] remains operative throughout. It is important in the CRZ, but is a minor H$_2$ removal process in the PDR.

Following \cite{Bialy2019} we write
\begin{equation}
\label{eq: R-}
    R_- = \eta k_1 x_{\rm e} \ \ \,
\end{equation}
where $x_{\rm e}\equiv n_e/n$ is the electron fraction, and $k_1$  
is the rate coefficient of the radiative attachment reaction [R1]. 
The branching ratio
\begin{equation}
\label{eq: eta}
    \eta = \frac{k_2 x_{\rm HI}}{k_2 x_{\rm HI}  +  k_3x_{{\rm H^+}}  +  D_-/n} 
\end{equation}
is the fraction of radiative attachments that are followed by associative detachment [R2] 
rather than mutual neutralization [R3] 
or photodetachment [R4]. For $\eta=1$ all radiative attachments result in H$_2$ formation.  

Throughout the PDR and into the CRZ, including where the gas is molecular, the electron and proton fractions are equal, i.e.~the positive charge is carried mainly by protons. 
This is because the positive molecular ions produced by CR ionization of H$_2$ [R8] are rapidly removed by abstraction [R10], followed by rapid dissociative recombination [R13], and [R14]. The fraction of H$_2$ ionizations leading to H$^+$ [R15] is small, and CR ionization of atomic hydrogen [R7] is the dominant source of electrons throughout.
With the further assumption that the gas is neutral, i.e.~$x_{\rm e}\ll 1$, (and with the neglect of helium ionization), we have that
\begin{equation}
\label{eq: xe}
    x_{\rm e} = \biggl(\frac{0.46\zeta} {n\alpha_{\rm B}}x_{\rm HI}\biggr)^{1/2} = \ 2.4\times 10^{-6}\bigl(\frac{\zeta_{-16}}{n_6}\bigr)^{1/2}T_2^{3/8}x_{\rm HI}^{1/2} \ \ \ ,
\end{equation}
where $\alpha_{\rm B}$
is the electron-proton recombinaton rate coefficient, $n_6=n/(10^6 \ {\rm cm^3})$, and $T_2= T/100 {\rm K}$.  In this approximation, $x_{{\rm H^+}}=x_{\rm e}$. 

The branching ratio $\eta$ becomes small ($\lesssim 0.5$) if the proton fraction
$x_{\rm H^+} \gtrsim (k_2/k_3)x_{\rm HI}$, or if $D_-\gtrsim k_2nx_{\rm HI}$.
For atomic gas ($x_{\rm HI}=1$),
$\eta<0.5$ for $x_{\rm H^+}>0.03$. This requires $\zeta_{-16}/n_6>2.21\times 10^8$ as given by Eq.(\ref{eq: xe})  (for $T=100$~K). For the $T=10^5$~K spectrum we assume in our computations, $\eta < 0.5$ and photodetachment becomes important for $I_{\rm LW}/n_6>(I_{\rm LW}/n_6)_{\rm crit}=1.6\times 10^6$.
In Table \ref{table: D_} we list the critical UV-intensity to density ratios, $(I_{\rm LW}/n_6)_{\rm crit}$, at which $\eta=0.5$ for the different spectral shapes. 

For comparison, in Table \ref{table: D_ metagalactic} we give the critical densities, $n_{6,crit}$, below which H$^-$ photodetachment becomes important for the (unnormalized) metagalactic fields at the various redshifts.  We also list these densities relative to the cosmic mean baryon densities $n_b$ at the different epochs.  Relative to the cosmic mean densities, photodetachment by the evolving metagalactic field is more effective at lower redshift.

For the parameter space we are considering it is safe to assume $\eta=1$, and we then have
\begin{equation}
\label{eq: Rminusb}
R_- = 4\times 10^{-22} \bigl(\frac{\zeta_{-16}}{n_6}\bigr)^{1/2}T_2^{1.02}
e^{-0.09/T_2}x_{\rm HI}^{1/2} \ \ {\rm cm^3 \ s^{-1} \ \ .} 
\end{equation}
Equation (\ref{eq: xe}) is valid for $\zeta_{-16}/n_6$ up to $\sim 1.7\times 10^9$ for which $x_{\rm e} \lesssim 0.1$ (for $x_{\rm HI}=1$). For this maximal ionization fraction (with the gas still mainly neutral) $R_-\approx 1.6\times 10^{-17}$ cm$^3$ s$^{-1}$ at 100~K.
Remarkably, this is comparable to the characteristic $R_{\rm dust}=3\times 10^{-17}$ cm$^3$ s$^{-1}$ for grain-surface catalysis of H$_2$ for typical Galactic dust-to-gas ratios, $Z^\prime_d=1$. However, 
for $\zeta_{-16}=1$ this requires extremely low density $n_6\sim 6\times 10^{-10}$ ($n=6\times 10^{-4}$~cm$^{-3}$). 
In general, $R_-\ll R_{\rm dust}$ 
 for typical ISM conditions.
 For $\zeta_{-16}=n_6=1$, $R_-=3.65\times 10^{-22}$~cm$^3$~s$^{-1}$ at 100 K (for $x_{\rm HI}=1$), and gas-phase formation becomes important for very small dust-to-gas ratios $Z^\prime_d\lesssim 1.3\times 10^{-5}(\zeta_{-16}/n_6)^{1/2}$. However, such low dust-to-gas ratios may not require correspondingly small metallicities because the dust abundances may decline sharply and non-linearly. For example, given the \cite{Remy-Ruyer2013} broken power-law representation for the gas-to-dust 
 ratio, Z$^\prime_d$, versus  overerall metallicity, Z$^\prime$, into the low metallicity regime \citep[see also][]{Bialy2019,Li2019}, gas phase H$_2$ formation dominates for $Z^\prime \lesssim 0.01(\zeta_{-16}/n_6)^{1/6}$ at 100 K. 
 Furthermore, $R_-$ increases with temperature, whereas $R_{\rm dust}$ probably declines \citep{Cuppen2010}, so the relative efficiencies of the gas-phase processes may grow in warmer gas \citep{Glover2003}.
 
In dusty clouds, electrons are provided by the photoionization of heavy elements, and dust neutralization processes may reduce the ionization fractions. As we discuss
 in Appendix A, heavy element ionization is negligible for $Z^\prime\lesssim 0.01(\zeta_{-16}/n_6)^{1/2}$, and dust grain neutralization is unimportant for $Z^\prime_d \lesssim 10^{-5}$. Coincidentally, this is the gas-to-dust ratio for which H$_2$ formation on dust may be ignored as well.
 
 Because $R_-$ is proportional to $(\zeta/n)^{1/2}$, whereas destruction is proportional to $(\zeta/n)$, the molecular fraction {\it decreases} with $\zeta/n$ in the CRZ as we discuss further below.
 For any $\zeta$, the effective rate coefficient $R_- \propto n^{-1/2}$, due to the lower electron fractions at larger densities. The {\it rate} of gas phase H$_2$ formation, $R_-n$ (s$^{-1}$), is therefore proportional to $n^{1/2}$. 
 
 Equation~(\ref{eq: xe}) may also be used to estimate the H$^-$ abundance.  Assuming formation via [R1] and removal by just [R2] 
we obtain the upper limit
\begin{align}
\label{eq: xHminus}
    x_{{H^-}} &\lesssim \ \frac{k_1}{k_2}x_{\rm e}  \nonumber \\
    &= 6.04\times 10^{-14} T_2 ^{1.41} e^{0.3/T_2} 
    \bigg{(}\frac{\zeta_{-16}}{n_6}\bigg{)}^{1/2} x_{\rm HI} ^{1/2}\ \ .
\end{align}
Although the H$^-$ anions are critical intermediaries for H$_2$ formation, their abundance remains extremely small for steady-state conditions.


\subsection{PDR}
In the PDR, Eq.(\ref{eq: form-des}) may be written as
\begin{equation}
\label{eq: fPDR}
    x_{\rm HI} = \frac{1}{2}\alpha f_{\rm shield}(N_{\rm H_2}) x_{\rm H_2}
\end{equation}
where 
\begin{align}
\label{eq: alpha}
    \alpha \equiv \frac{D_0}{R_-n} &= \frac{\sigma_d{\bar F}_{\nu,{\rm LW}}}{R_-n}  \\ \nonumber
    &= 1.45\times 10^5\bigl(\frac{I_{\rm LW}}{n_6}\bigr)\bigl(\frac{\zeta_{-16}}{n_6}\bigr)^{-1/2}T_2^{-1.02}x_{\rm HI}^{-1/2} \ \ \ .
\end{align}
The unattenuated photodissociation rate, $D_0$, depends on the far-UV field strength, $I_{\rm LW}$, so the dimensionless parameter $\alpha$ is a function of the two ratios $I_{\rm LW}/n$ and $\zeta/n$ (via $R_-$). In general the gas is atomic ($x_{\rm HI}=1$) at the unshielded cloud boundary, and the H$_2$ fraction is
\begin{equation}
\label{eq: x2}
    x_{\rm H_2} = \frac{2}{\alpha} = 1.38\times 10^{-5}\bigl(\frac{I_{\rm LW}}{n_6}\bigr)^{-1}\bigl(\frac{\zeta_{-16}}{n_6}\bigr)^{1/2} \ \ \ .
\end{equation}
For fiducial parameters, $I_{\rm LW}=\zeta_{-16}=n_6=1$, the molecular fraction $x_{\rm H_2}=1.38\times 10^{-5}$ at the cloud boundary, and a molecular self-shielding column of $\gtrsim 10^{21}$~cm$^{-2}$ for which $f_{\rm shield} \lesssim 10^{-5}$ is required to enable an H{\small I}-to-H$_2$ transition.

Equation (\ref{eq: fPDR}) may be written as the differential equation
\begin{equation}
    dN_{\rm HI} = \frac{1}{2}\alpha f_{\rm shield}(N_{\rm H_2}) dN_{\rm H_2}
\end{equation}
for the H{\small I} column as a function of the H$_2$ column. For a constant $\alpha$, independent of cloud depth, 
integrating gives
\begin{equation}
\label{eq: difff}
    N_{\rm HI}(N_{\rm H_2}) = \frac{\alpha}{2} \int_0^{N_{\rm H_2}}f_{\rm shield}(N_{\rm H_2}')dN_{\rm H_2}' = \alpha \frac{W_d(N_{\rm H_2})}{\sigma_d} \ \ \ ,
\end{equation}
where $\sigma_d$ is the line dissociation cross section (cm$^2$ Hz), and $W_d(N_{\rm H_2})$ is the dissociation bandwidth, as given by Equation~(\ref{eq: Wfit}). In pulling $\alpha$ out of the integral we are assuming (a) constant gas density for the given $I_{\rm LW}$ and $\zeta_{-16}$, (b) that $x_{\rm HI}^{1/2}\approx 1$, i.e.~the gas is largely atomic, where photodissociation dominates the production of the H{\small I} column, and (c) no CR attenuation so that $\zeta_{-16}$ is constant. Equation (\ref{eq: difff}) can also be written as \begin{equation}
    N_{\rm HI}(N_{\rm H_2}) = \frac{1}{2}\frac{{\bar F_{{\nu},{\rm LW}}}W_d(N_{\rm H_2})}{R_-n} \ \ \ \ \ .
\end{equation}
The numerator, ${\bar F}_{\nu, {\rm LW}}W_d(N_{\rm H_2})/2,$ is the radiation flux absorbed in dissociations up to a molecular column $N_{\rm H_2}$, and this is equal to the product $R_-nN_{\rm HI}(N_{_{\rm H_2}})$, i.e.~the rate at which molecules are produced within $N_{\rm H_2}$.


For sufficiently large $N_{\rm H_2}$ ($\gtrsim 10^{22}$~cm$^{-2}$) the H$_2$ absorption lines fully overlap, and all of the LW band photons are absorbed by the H$_2$. The total H{\small I} column maintained by photodissociation by LW band photons is then
\begin{align}
\label{eq: NUVtot}
    N_{{_{\rm HI,{\rm tot}}}}^{\rm LW} &= \frac{\alpha}{2}\frac{W_{d,{\rm tot}}}{\sigma_d} = \frac{1}{2}\frac{{\bar F_{\nu,{\rm LW}}}W_{d,{\rm tot}}}{R_-n} \\ 
    \nonumber
    & = 2.7\times 10^{21}
    \bigl(\frac{I_{\rm LW}}{n_6}\bigr) \bigl(\frac{\zeta_{-16}}{n_6}\bigr)^{-0.5} \ \ \ \ {\rm cm}^{-2} \ \ \ .
\end{align}
In the third equality we have set $T_2=1$, and $x_{\rm HI}=1$ in the expression for $R_-$ with the assumption that the photodissociated H{\small I} is built up in an extended fully atomic PDR and the transition from H{\small I} to H$_2$ is sharp. This assumption is valid if the transition occurs as the absorption lines overlap, in which case the dissociation rates are reduced exponentially leading to sharp transitions. Setting $x_{\rm HI}=1$ is more approximate when the transition points occur before the lines overlap in which case further LW absorptions out of the damping wings enable continued gradual growth of the photodissociated H{\small I} columns in the molecular zones. Sharp transitions induced by line-overlap occur when $N_{{_{\rm HI,{\rm tot}}}}^{\rm LW} \gtrsim 10^{22}$~cm$^{-2}$. 

We select fiducial parameters, $I_{\rm LW}=\zeta_{-16}=n_6=1$ (or $I_{\rm LW}/n_6=\zeta_{-16}/n_6=1$) such that the resulting H{\small I} photodissociation column, $N_{{_{\rm HI,{\rm tot}}}}^{\rm LW}=2.7\times 10^{21}$~cm$^{-2}$, is comparable to the H$_2$ shielding column $\sim 10^{22}$ required for line overlap.




It is of interest to compare Eq.~(\ref{eq: NUVtot}) to the
\citetalias{Sternberg2014} formula for the total H{\small I} photodissociation column in dusty clouds, where molecule formation is via grain catalysis, and opacity is provided by FUV dust absorption in addition to H$_2$ self-shielding. The \citetalias{Sternberg2014} formula (for beamed fields) is
\begin{equation}
\label{eq: NUVtot_dust}
    N_{{_{\rm HI,{\rm tot}}}}^{\rm LW} = \frac{1}{\sigma_g}{\rm ln}\bigl[\frac{1}{2}\frac{\sigma_g {\bar F}_\nu W_{g,{\rm tot}}}{R_{\rm dust}n}+1\bigr] \ \ \ ,
\end{equation}
where $\sigma_g$ is the dust absorption cross section, $R_{\rm dust}$ is the grain-surface H$_2$ formation rate coefficient, and $W_{g,{\rm tot}}$ is the ``H$_2$ dust limited dissociation bandwidth" (see \citetalias{Sternberg2014}).  For vanishing dust abundance, $\sigma_g\rightarrow 0$, $W_{g,{\rm tot}}\rightarrow W_{d,{\rm tot}}$, and $R_{\rm dust}$ may be replaced with the gas-phase coefficient $R_-$. Expanding the logarithm, $\sigma_g$ cancels out, and this yields our Eq.(\ref{eq: NUVtot}) for dust-free conditions. 
We further discuss Eq.~(\ref{eq: NUVtot}) versus (\ref{eq: NUVtot_dust}) in Appendix B.

\subsection{CRZ}
The transition from the outer PDR to the inner cosmic-ray zone (CRZ) occurs 
as self-shielding reduces the H$_2$ photodissociation rate below the CR destruction rate. In the CRZ, H$_2$ formation continues to be driven by CR ionization. Removal is by a combination of CR ionization and dissociation.

Moving from the PDR to CRZ may, or may not, give rise to an atomic to molecular transition, depending on $\zeta/n$, as follows.  In the CRZ, Eq.(\ref{eq: form-des}) may be written as
\begin{equation}
\label{eq: crfd}
    R_-nx_{\rm HI} = [0.96 f_d(1+f_3) + 0.74]\zeta x_{\rm H_2} \ \ \ .
\end{equation}
In analogy to Eq.(\ref{eq: fPDR}) for the PDR, this may be written as
\begin{equation}
    x_{\rm HI}=\beta x_{\rm H_2}
\end{equation}
where implicitly
\begin{align}
\label{eq: beta}
    \beta  &\equiv  [ 0.96f_d(1+f_3)+0.74] \frac{\zeta}{R_-n} \\
    \nonumber
     &= [0.96(1+f_3)\frac{1}{1+k_6x_{\rm HI}/k_{10}x_{\rm H_2}}+0.74] \frac{\zeta}{R_-n} \\
     \nonumber
     &=0.25\times [\frac{1.58}{1+0.31x_{\rm HI}/x_{\rm H_2}}+0.74]\bigl(\frac{\zeta_{-16}}{n_6}\bigr)^{1/2} \\
     \nonumber
     &\times
     T_2^{-1.02}e^{0.092/T_2}x_{\rm HI}^{-1/2}
     \ \ \ .
\end{align}
Here we have used Eqs.~(\ref{eq: fd}), (\ref{eq: R-}), and (\ref{eq: xe}), with $\eta=1$. It follows that the atomic and molecular mass fractions are equal, ($x_{\rm HI}=2x_{\rm H_2}$, or $\beta=2$) for the critical ratio\footnote{
This expression for the critical ionization to density ratio is more accurate than Eq.~21 in \cite{Bialy2015a} for which  $x_{\rm HI}=2x_{\rm H_2}$ at $\zeta_{-16}/n_6=11.4$. In that paper we made the approximations $f_d=1$, and $1+f_3=2$, and we set the cosmic-ray dissociation rate to $0.1\zeta$ (as incorrectly listed in UMIST12) rather than to $0.7\zeta$ as here.
}
\begin{equation}
\label{eq: crit}
\biggl(\frac{\zeta_{-16}}{n_6}\biggr)_{\rm crit} = 10.88\times  T_2 ^{2.03} e^{-0.18/T_2} \ = \ 9.02 \ \ \ ,
\end{equation}
Equation (\ref{eq: crfd}) is the algebraic equation

\begin{align}
\label{eq: crbal}
& 3.96 \biggl(\frac{\zeta_{-16}}{n_6}\biggr)^{1/2}x_{\rm HI}^{3/2}T_2^{1.02}e^{-0.092/T_2}  =
\\& \nonumber [0.96(1+f_3)\frac{1}{1+k_6x_{\rm HI}/k_{10}x_{\rm H_2}}+ 0.74]\frac{\zeta_{-16}}{n_6}x_{\rm H_2}
\end{align}
The left-hand side is the formation rate per unit volume, and is proportional to $(\zeta/n)^{1/2}$. The right-hand side is the destruction rate per unit volume, and is proportional to $\zeta/n$. For $\zeta/n$ less than the critical value given by Eq.(\ref{eq: crit}) the gas becomes molecular in the CRZ, and an H{\small I}-to-H$_2$ transition then occurs in moving from the PDR to the CRZ.  For larger $\zeta/n$ the gas remains atomic and an H{\small I}-to-H$_2$ transition does not occur.

In Figure~4 we plot the atomic and molecular fractions as functions of $\zeta_{-16}/n_6$ as given by the solution to Eq.~(\ref{eq: crbal}) (dashed curves) compared to the numerical solutions (solid curves) of the full rate equations within the CRZ. The agreement is excellent. The vertical dotted line indicates the critical ratio of 9.02 as given by Eq.~(\ref{eq: crit}).  This is close to our numerically computed value of 11.84 indicated by the vertical solid line.

\subsection{Analytic Profiles}

\begin{figure}
	\centering
	\includegraphics[width=1\columnwidth]{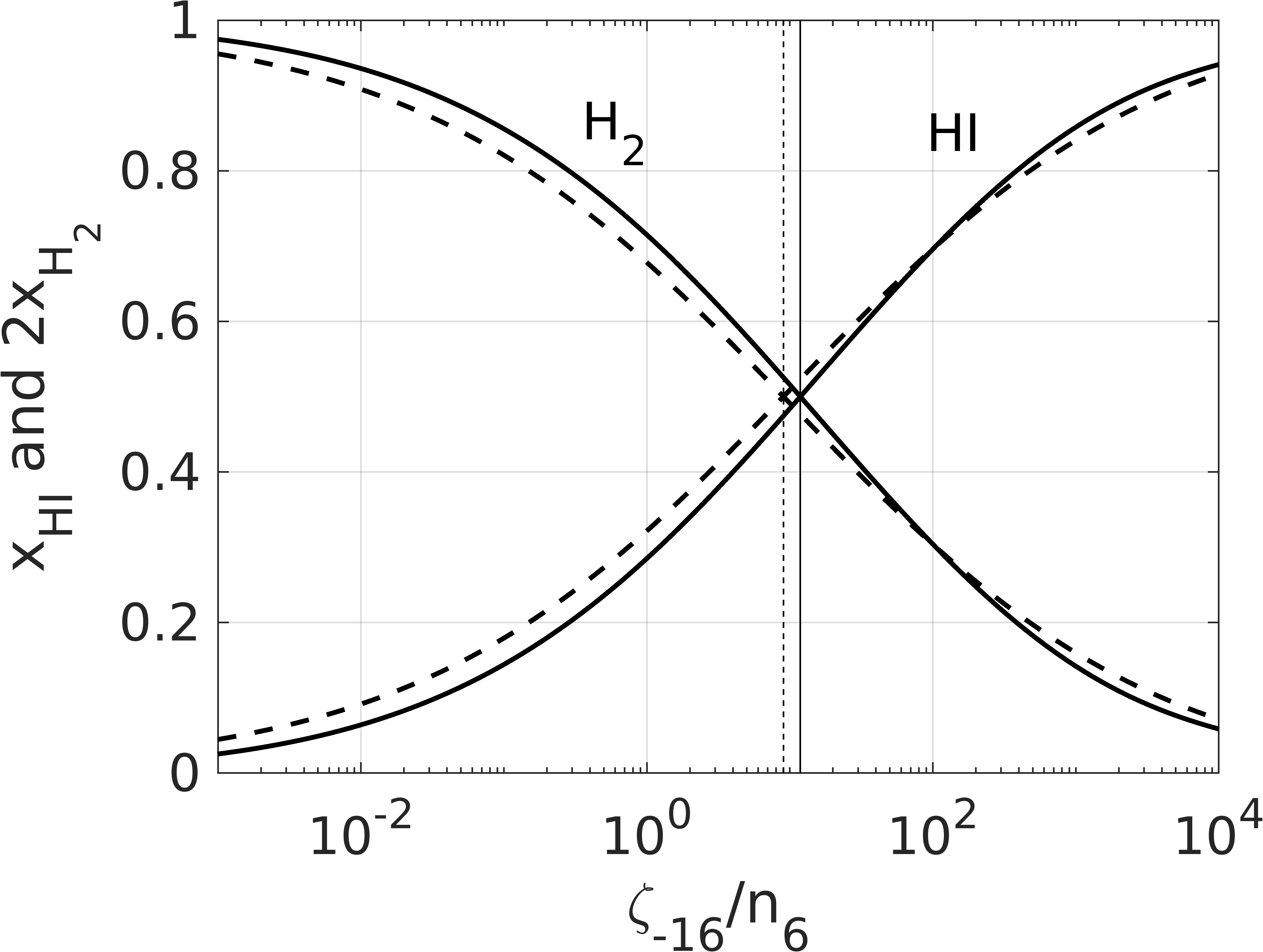} 
	\caption{
Atomic and molecular gas mass fractions, $x_{\rm HI}$ and $2x_{\rm H_2}$, as functions of $\zeta_{-16}/n_6$ (for $T_2=1$) as given by solutions to Eq.~(\ref{eq: crbal}) (dashed curves), and the numerical solutions to the full set of rate equations (solid curves). The vertical dashed and solid lines indicate the critical value as given by the analytic Eq.~(\ref{eq: crit}) and the numerical solutions respectively.
		}
		\label{fig: crz x1x2}
\end{figure}

Our procedure to generate analytic profiles is based on the formation-destruction equation
\begin{equation}
\label{eq: a1}
    x_{\rm HI} = [\frac{1}{2}\alpha f_{\rm shield}(N_{\rm H_2}) + \beta]x_{\rm H_2} \ \ \ ,
\end{equation}
and mass conservation
\begin{equation}
\label{eq: a2}
    x_{\rm HI}+2x_{\rm H_2}=1 \ \ \ .
\end{equation}
With our expressions for $\alpha$ and $\beta$ (Eqs.~[\ref{eq: alpha}] and [\ref{eq: beta}]) 
the local abundances $x_{\rm HI}$ and $x_{\rm H_2}$ may be computed as functions of the molecular column density $N_{\rm H_2}$, for any given $\zeta/n$ and $I_{\rm LW}/n$.
The parameters $\alpha$ and $\beta$ themselves depend on $x_{\rm HI}$ and $x_{\rm H_2}$. We integrate Eq.(\ref{eq: a1}) assuming that they are constant, writing them as ${\bar \alpha}$ and ${\bar \beta}$ independent of cloud depth, and assuming no CR attenuation. For ${\bar \alpha}$, we set $x_{\rm HI}=1$ in Eq.(\ref{eq: alpha}). For ${\bar \beta}$, and for any  ${\zeta_{-16}/n_6}$, we set $f_d$ as given by Eq.(\ref{eq: fd}) for the constant (asymptotic) $x_{\rm HI}/x_{\rm H_2}$ ratio in the optically thick CRZ.  We have computed $f_d$ and ${\bar\beta}$, using our full numerical solutions for $x_{\rm HI}/x_{\rm H_2}$ in the CRZ shown in Figure \ref{fig: crz x1x2}.  We find that for 50~K~$<T<$~10$^3$~K the power-law fit
\begin{equation}
    {\bar{\beta}}=0.75\times \bigl(\frac{\zeta_{-16}}{n_6}\bigr)^{0.37} T_2 ^{-0.79}
    \label{eq: betafit}
\end{equation}
is accurate to within 11\% for $0.1<\zeta_{-16}/n_6<10$ and 34\% for $0.01<\zeta_{-16}/n_6<100$.
With these approximations,
\begin{equation}
\label{eq: a3}
    N_{\rm HI}(N_{\rm H_2}) = \frac{{\bar \alpha}}{2}\frac{W_d(N_{\rm H_2})}{\sigma_d} + {\bar\beta} N_{\rm H_2} \ \ \ .
\end{equation}
The first term on the right is the atomic column built up in the PDR, the second term is the column built up in the CRZ.

The depth-dependent atomic and molecular densities may then be constructed in the following simple procedure:
\begin{enumerate}
    \item Select the ratios $I_{\rm LW}/n$, and $\zeta/n$.
    \item Compute ${\bar \alpha}$ using Eq.~(\ref{eq: alpha}) assuming $x_{\rm HI}=1$.
    \item Compute ${\bar \beta}$ using Eq.~(\ref{eq: betafit}) given the assumed $\zeta/n$.
    \item Solve Eqs.~(\ref{eq: a1}) and (\ref{eq: a2}) for $x_{\rm HI}$ and $x_{\rm H_2}$, using ${\bar \alpha}$ and ${\bar \beta}$, given any $N_{\rm H_2}$ and the shielding function Eq.~(\ref{eq: shield}).
    \item Compute $N_{\rm HI}(N_{\rm H_2})$ using Eq.~(\ref{eq: difff}) and with Eq.~(\ref{eq: Wfit}) for $W_d(N_{\rm H_2})$.
    \item Plot the density profiles $x_{\rm HI}(N)$ and $x_{\rm H_2}(N)$, with $N=N_{\rm HI}+2N_{\rm H_2}$.
\end{enumerate}
As we demonstrate below, this procedure provides an accurate representation for the density profiles and H{\small I}-to-H$_2$ transition points for models without CR attenuation, compared to full numerical solutions to the depth-dependent rate equations for the hydrogen/helium network.

\subsection{Timescales}

We have been assuming steady-state conditions for the atomic and molecular densities.  The time-scales required to reach a steady-state equilibrium may be estimated by considering the time-dependent H$_2$ formation-destruction equation \citep{Goldsmith2007, Liszt2007}. In our problem this may be written as 
\begin{equation}
    \frac{dx_{\rm H_2}}{d\tau} = 1 - [2 + \frac{1}{2}\alpha f_{\rm shield}(N_{\rm H_2}) + \beta]x_{\rm H_2}
\end{equation}
where here the dimensionless time $d\tau\equiv R_-ndt$. Making the assumption that $\alpha$ and $\beta$ are independent of $x_{\rm HI}$ and $x_{\rm H_2}$ as the system evolves, 
e.g.~by setting $\alpha={\bar \alpha}$ and $\beta={\bar \beta}$, the solution is
\begin{equation}
    x_{\rm H_2} = 1/a + [x_{\rm H_2}(0)-1/a)]e^{-a\tau}
\end{equation}
where $x_{\rm H_2}(0)$ is the initial molecular fraction, and
\begin{equation*}
 a\equiv 2 + \frac{1}{2}{\bar \alpha}f_{\rm shield}(N_{\rm H_2}) + {\bar \beta} \ \ \ .
\end{equation*}
The  H{\small I}-to-H$_2$ equilibration time scale for any shielding column $N_{\rm H_2}$ is then
\begin{equation}
    \tau_{\rm eq} = \frac{1}{a} \ \ \ ,
\end{equation}
or
\begin{equation}
\label{eq: teq}
    t_{\rm eq} = \frac{t_{\rm H_2}}{a}=\frac{1}{R_-n}\times \frac{1}{2 + \frac{1}{2}{\bar \alpha} f_{\rm shield}+{\bar \beta}} \ \ \ ,
\end{equation}
 where the H$_2$ formation time
\begin{equation}
t_{\rm H_2} \equiv \frac{1}{R_-n}
    = \frac{8.0\times 10^7}{n_6} \biggl(\frac{\zeta_{-16}}{n_6}\biggr)^{-1/2}x_{\rm HI}^{-1/2}T_2^{-1.02} \ \  {\rm yr}
\end{equation}
using Eq.(\ref{eq: Rminusb}) for $R_-$.

In general, $\alpha\gg 1$ and $\beta$ is of order unity (see Eqs.~[\ref{eq: alpha}] and [\ref{eq: beta}]). Therefore, in the outer PDR where $f_{\rm shield}\rightarrow 1$, $a\gg 1$, and the equilibration time-scale is short, with
\begin{equation}
    t_{\rm eq} \approx \frac{2}{D_0}=\frac{1.1\times 10^3}{I_{\rm LW}} \ \  {\rm yr} \ \ \ .
\end{equation}
In this limit the equilibration time is the dissociation time, 
independent of the gas density.  If the gas is initially molecular ($x_{\rm H_2}(0)=1/2$) photodissociation to the equilibrium atomic state is rapid. Or, if initially fully atomic ($x_{\rm H_2}(0)=0$) the small equilibrium molecular abundance is reached within a short time.
In the inner shielded CRZ, where $f_{\rm shield}\rightarrow 0$, and with $\beta \lesssim 1$, we have $a\approx 1/2$, and
\begin{equation}
    t_{\rm eq} \approx \frac{1}{2R_-n} = \frac{3.96\times 10^7}{n_6} \biggl(\frac{\zeta_{-16}}{n_6}\biggr)^{-1/2}x_{\rm HI}^{-1/2}T_2^{-1.02} \ \  {\rm yr} \ \ \ .
\end{equation}
The equilibration time is then long and comparable to the H$_2$ formation time. In this limit, $t_{\rm eq}$ is independent of the photodissociation rate, is inversely proportional to the gas density for fixed $\zeta/n$, and  decreases as $\zeta^{-1/2}$ for fixed density.

For our fiducial parameters, $I_{\rm LW}=\zeta_{-16}=n_6=T_2=1$, we have that ${\bar \alpha}=1.45\times 10^5$, and ${\bar \beta}\approx 0.75$. Thus, in the unshielded PDR, $t_{\rm eq}\approx 1.1\times 10^3$~yr. In the CRZ, $t_{\rm eq}\approx 4.3\times 10^7$ yr. In Fig. \ref{fig: timescale grid}, we show computations of the depth-dependent time-scales, short to long, and shown as $n_6\times t_{eq}$, for a range of $I_{\rm LW}/n_6$ and $\zeta_{-16}/n_6$. We discuss these results as part of our parameter study in \S 4.4.

\section{Numerical Models}

We now present solutions for the atomic and molecular abundance profiles, and the H{\small I}-to-H$_2$ transition points, obtained by solving the rate equations (Eq.~[\ref{eq: rateeq}]) for our hydrogen-helium chemical network (Figure \ref{fig: chem_network} and Table \ref{table: chem}). We refer to these solutions as our ``numerical models". We write down the rate equations, and then present our fiducial case and model grid. We compare the results to the ``analytic" description we discussed in \S 3.
\begin{figure}
	\centering
	\includegraphics[width=1\columnwidth]{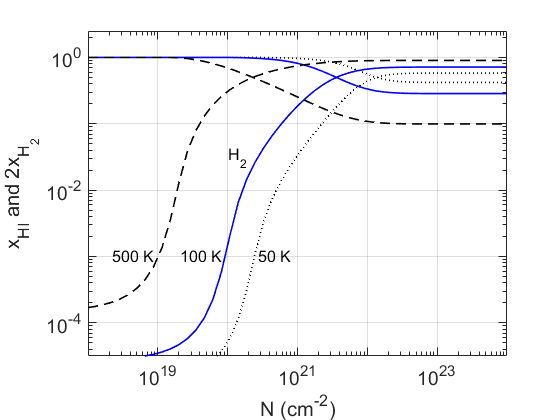} 
	\caption{
The HI and H$_2$ mass fractions, $x_{\rm HI}$ and $2x_{\rm H_2}$, for our fiducial parameters $I_{\rm LW}/n_6=1$ and $\zeta_{-16}/n_6=1$, for gas temperatures $T$=50~K (dotted), 100~K (solid blue), and 500~K (dashed), and no CR attenuation.
		}
		\label{fig: temp grid}
\end{figure}

\begin{figure*}
    
	\centering
	\includegraphics[width=0.8\textwidth]{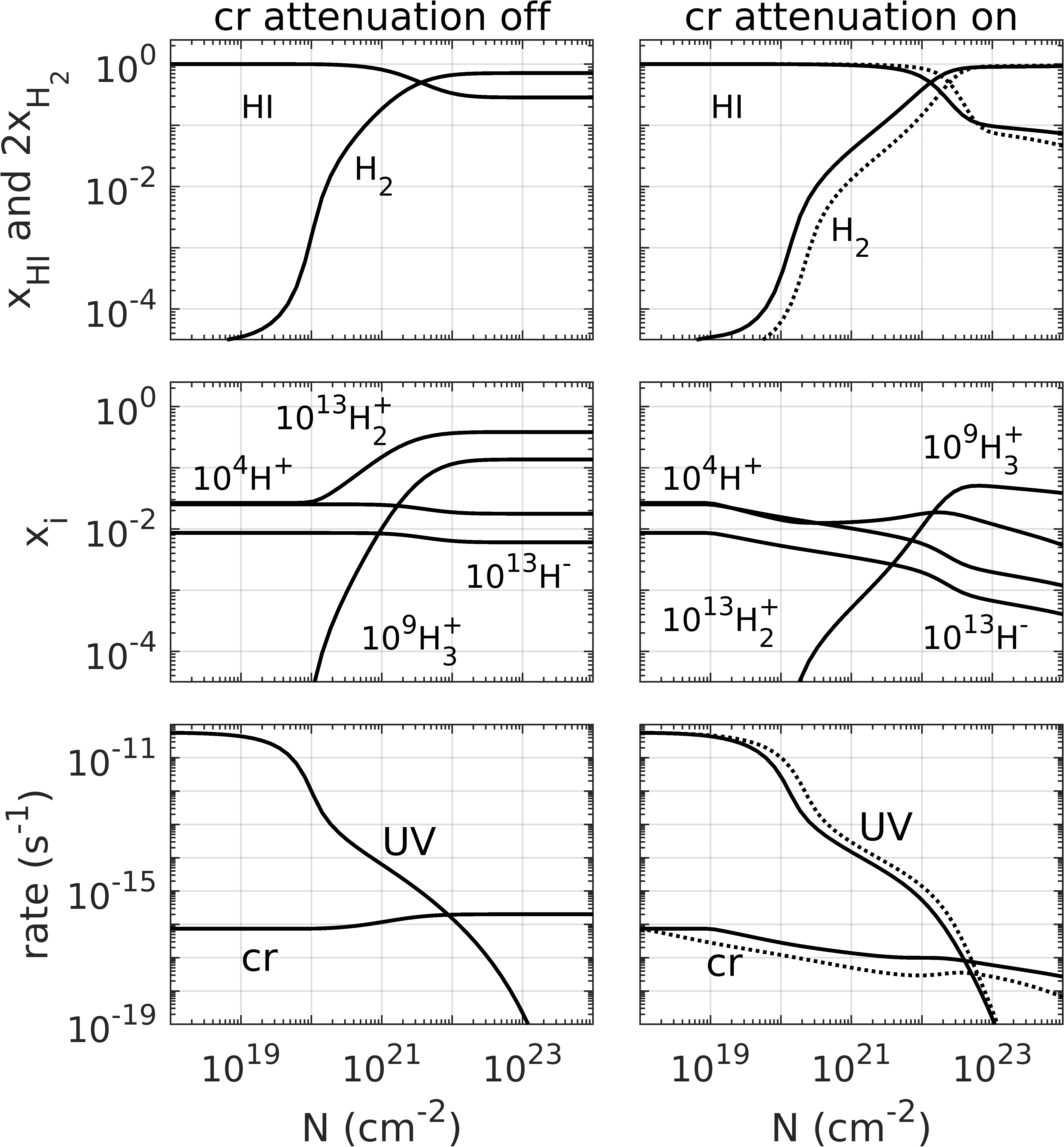} 
	\caption{Fiducial models for $T_2=1$, without (left panels) and with (right panels) CR attenuation. The upper row shows the HI and H$_2$ mass fractions. For the models with CR attenuation two sets of curves are shown, for magnetic field orientations $\mu=1$ (solid) and 0.1 (dotted).  Middle rows show the fractional abundances of electrons, H$^+$, H$^-$ and H$_3^+$ (for $\mu=1$). The bottom row shows the FUV and CR destruction rates for $\mu=1$ (solid) and $\mu=0.1$ (dotted).
		}
		\label{fig: fiducial_case_2}
\end{figure*}

\begin{figure*}
    
	\centering
	\includegraphics[width=1\textwidth]{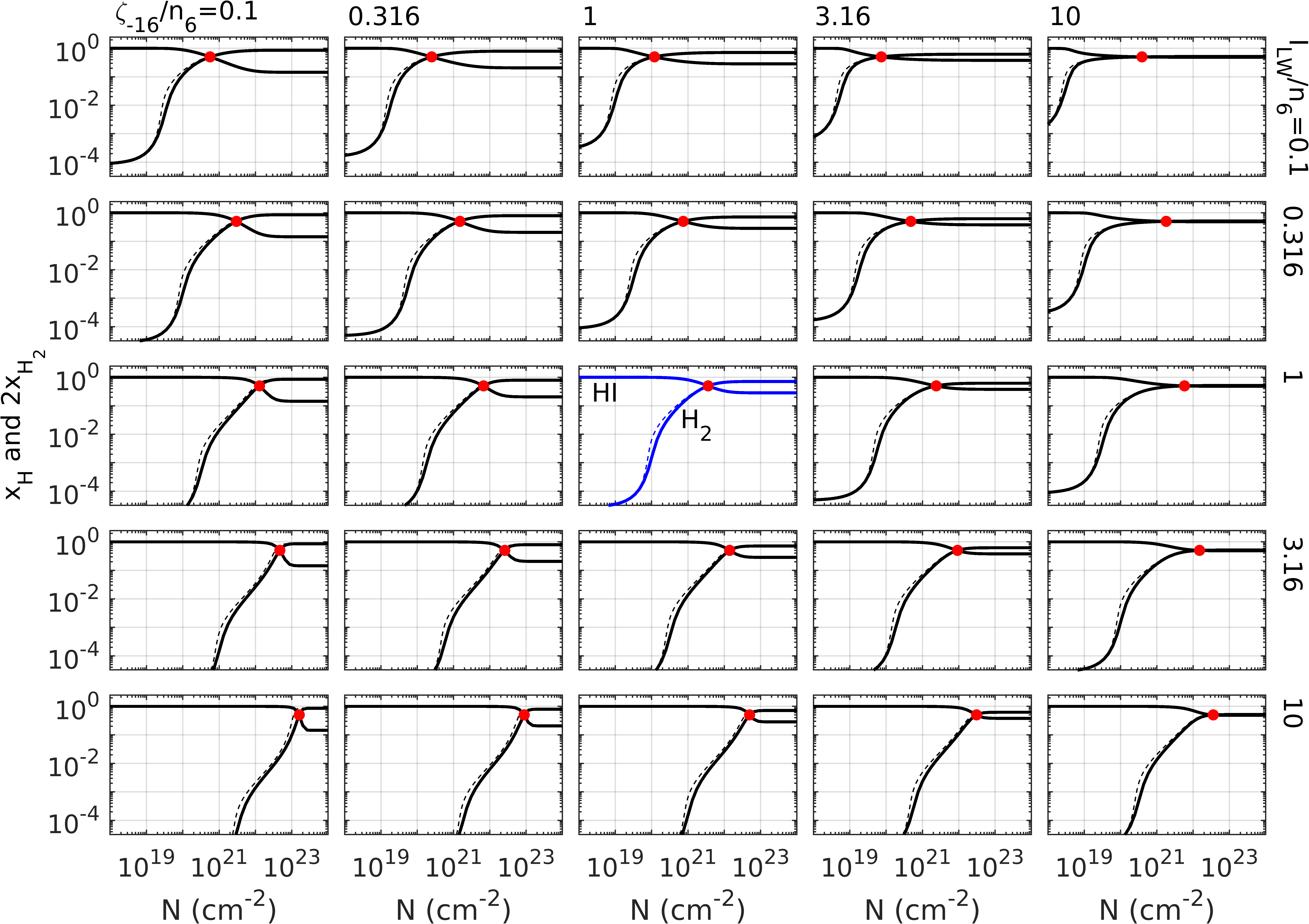} 
	\caption{
Grid of HI and H$_2$ mass fraction profiles for models without CR attenuation, for $I_{\rm LW}/n_6$ from 0.1 to 10, and $\zeta_{-16}/n_6$ also from 0.1 to 10. The dashed curves are the H$_2$ mass fractions calculated using the analytic procedure(\S 3.4). The red markers indicate the HI-to-H$_2$ transition points where $x_{\rm HI}=2x_{\rm H_2}$.
		}
		\label{fig: parameter_grid}
\end{figure*}

\begin{figure*}
    
	\centering
	\includegraphics[width=1\textwidth]{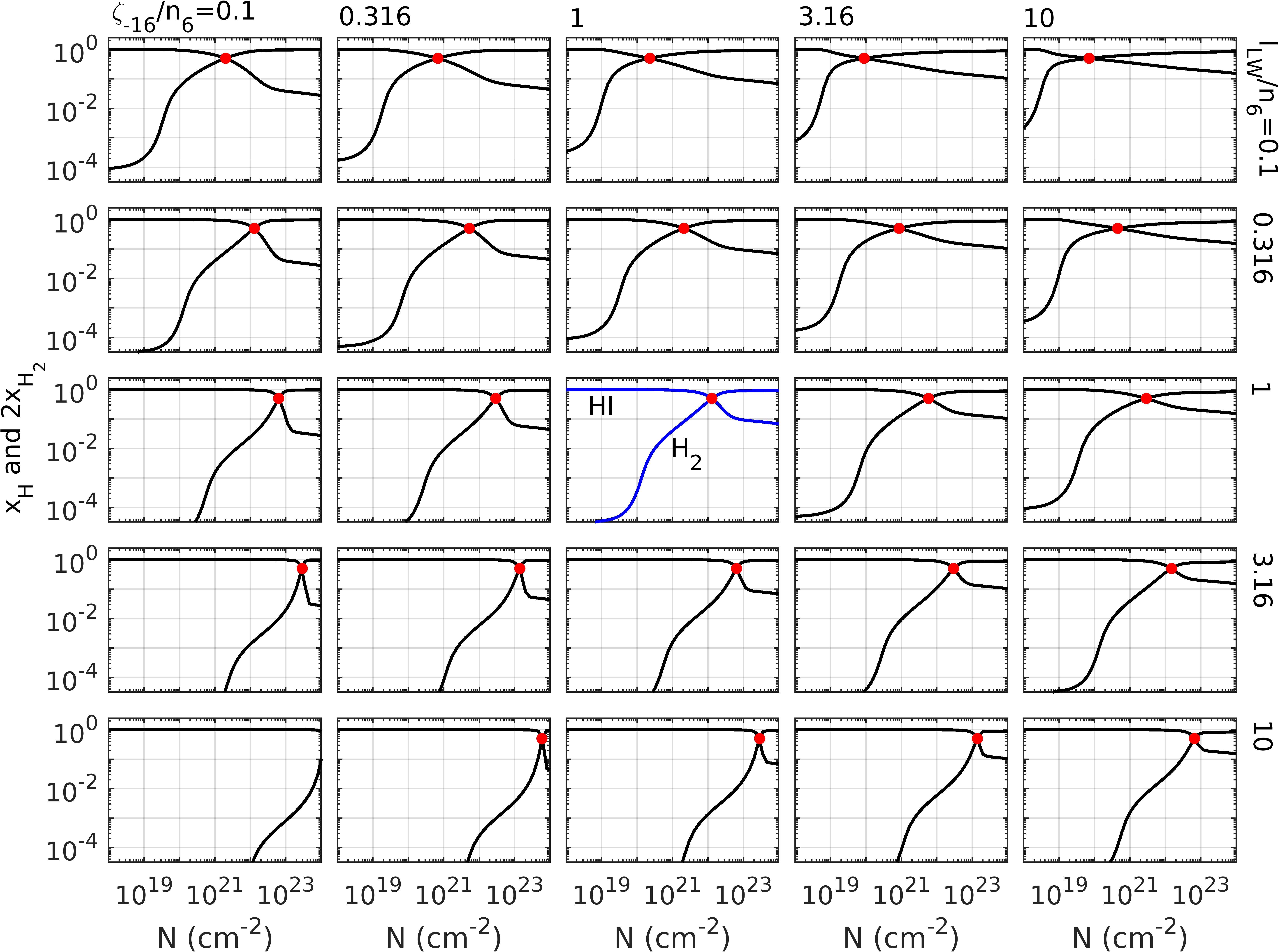} 
	\caption{
	Grid of HI and H$_2$ mass fraction profiles for models including CR attenuation, for $I_{\rm LW}/n_6$ from 0.1 to 10, and $\zeta_{-16}/n_6$ also from 0.1 to 10. The dashed curves are the H$_2$ mass fractions calculated using the analytic procedure(\S 3.4). The red markers indicate the HI-to-H$_2$ transition points where $x_{\rm HI}=2x_{\rm H_2}$. 
}
		\label{fig: parameter_grid_att}
\end{figure*}
\subsection{Rate Equations}
We solve the rate equations for the steady-state abundances, $x_i\equiv n_i/n$ of the atomic and molecular hydrogen/helium species in our chemical network described in \S 2. Here, $n_i$ is the density of species $i$, and $n$ is the total hydrogen gas density. The rate equations are
\begin{align}
\label{eq: rateeq}
&\sum_{jl} k_{ijl}(T) \, x_j \, x_l \, + \,  \frac{\zeta}{n}\sum_{j} a_{ij}  \, x_j\, + \,  \frac{I_{\rm LW}}{n} \sum_{j} b_{ij} \, x_j \nonumber\\ 
& = x_i \left( \sum_{jl} k_{jil}(T) \, x_l + \, \frac{\zeta}{n} \sum_{j} a_{ji} \,  + \, \frac{I_{\rm LW}}{n} \sum_{j} b_{ji} \right) \, .
\end{align}
The $k_{ijl}$ are the temperature dependent rate coefficients (cm$^3$ s$^{-1}$) for two-body reactions of species $j$ and $l$ that lead to the formation of $i$. We use the rate-coefficients as given in the UMIST2012 database \citep{McElroy2013}, and as updated by \cite{Neufeld2020}. We list these in Table \ref{table: chem}. The $a_{ij}$ are the direct cosmic-ray ionization and dissociation factors for the five hydrogen/helium CR reactions in our network as listed in Table \ref{table: cr},  multiplied by the CR attenuation factor. The $b_{ij}$ are the normalized photorates, for H$^-$ photodetachment, H$_2^+$ and HeH$^+$ photodissociation, and local H$_2$ photodissociation including self-shielding. The incident LW field intensity is specified by $I_{\rm LW}$, and $\zeta$ is the cosmic-ray ionization rate at the cloud edge.

Mass conservation is
\begin{equation}
    \sum_i \alpha_{im} x_i = X_m
\end{equation}
where $\alpha_{im}$ is the number of atoms of element $m$ (hydrogen or helium) contained in species $i$. The hydrogen abundance $X_{\rm H}=1$ by definition. For helium we assume a cosmic value $X_{\rm He}=0.1$. Charge conservation is the (dependent) equation
\begin{equation}
    \sum q_ix_i = 0
\end{equation}
where $q_i$ is the net charge of species $i$.

For any temperature $T$, and for a given H$_2$ column density $N_{\rm H_2}$, the solutions to the rate equations depend on the two ratios $I_{\rm LW}/n$ and $\zeta/n$. $N_{\rm H_2}$ determines the local photodissociation rate via the self-shielding factor. If CR attenuation is included the local solutions also depend on the total gas column $N=N_{\rm HI}+2N_{\rm H_2}$, but $N_{\rm H_2}$ may still be used as the single independent depth variable. Thus for isothermal conditions the depth-dependent abundance profiles are determined by the two ratios ${I_{\rm LW}/n}$ and $\zeta/n$ at the cloud boundaries.
Equations (\ref{eq: rateeq}) are non-linear, and we solve them numerically by integrating the time dependent differential rate equations to equilibrium, or via Newton iteration of the steady-state algebraic equations. We compute the species abundances, including $x_{\rm HI}$ and $x_{\rm H_2}$, as functions of cloud depth, integrating inward for $N_{\rm H_2}$ and $N_{\rm HI}$, given the initial conditions $N_{\rm HI}=N_{\rm H_2}=0$ at the outer boundary. We have verified (as a numerical consistency check) that for fixed temperature the solutions depend only on the ratios $I_{\rm LW}/n$ and $\zeta/n$, for any FUV intensity, cosmic-ray ionization rate, and gas density.

\subsection{Fiducial Case}

For our fiduical parameters we set $I_{\rm LW}/n_6=1$ and $\zeta_{-16}/n_6=1$. 
In Figure \ref{fig: temp grid} we show the resulting H{\small I} and H$_2$ mass fractions, $x_{\rm HI}$ and $2x_{\rm H_2}$, as functions of total gas column $N$ (cm$^{-2}$) for gas temperatures $T=50$, 100, and 500~K. In these computations we do not include any cosmic-ray attenuation. With these choices, (a) the gas is predominantly atomic in the outer PDR (Eq.~[\ref{eq: x2}]), 
(b) the gas is primarily molecular in the CRZ (Eq.~[\ref{eq: crit}]), and (c) the transition points occur at total gas columns $\lesssim 10^{22}$~cm$^{-2}$ (Eq.~[\ref{eq: NUVtot}]). 
 The effective gas-phase formation rate coefficient varies linearly with $T$ (Eq.~[\ref{eq: Rminusb}]), so as the temperature increases the transition points occur at smaller gas columns, and are at $1.4 \times 10^{22}$, $3.7 \times 10^{21}$, and $2.3 \times 10^{20}$~cm$^{-2}$, for $T=50$, 100, and 500~K. In the CRZ, the molecular fractions are 0.58, 0.71, and 0.90. It is evident that higher temperatures induce stronger H{\small I}-to-H$_2$ transitions.

In Figure~\ref{fig: fiducial_case_2} we present more details for the 
$T=100$~K case (lefthand panels), and we compare to models that include cosmic-ray attenuation (righthand panels). We set $\zeta_{-16}/n_6=1$ at the cloud surface, and assume the $q=0.385$ power-law (Eq.[\ref{eq: Pfit}]) for $\zeta(N)$ for the \citet{Padovani2018a} model-${\cal H}$ particle energy distribution. In the upper righthand panel, the solid curves are the H{\small I} and H$_2$ mass fractions assuming any magnetic field is normal to the cloud surface ($\mu=1$ in Eq.[\ref{eq: Pfit}])]. The dashed curves are for a highly inclined magnetic field ($\mu\equiv cos\theta=0.1$).

At the unshielded cloud edge, the molecular fraction $x_{\rm H_2}=1.3\times 10^{-5}$ (Eq.[\ref{eq: x2}]) for all three models. Without CR attenuation (upper left panel), the H{\small I}-to-H$_2$ transition point ($x_{\rm HI}=1/2$) occurs at $N=4\times 10^{21}$~cm$^{-2}$, where $N_{\rm HI}=2.9\times 10^{21}$ and $N_{\rm H_2}=5.7\times 10^{20}$~cm$^{-2}$. 
The H{\small I} column at the transition point is consistent with Eq.(\ref{eq: NUVtot}) for the total H{\small I} column produced by photodissociation. An H{\small I}-to-H$_2$ transition is expected because $\zeta_{-16}/n_6<11.87$ (see Fig.~\ref{fig: crz x1x2}). In the CRZ, the atomic and molecular mass fractions are constant, with $x_{\rm HI}=0.28$, and $2x_{\rm H_2}=0.72$.

With CR attenuation the transition point is shifted to slightly larger gas columns of $1.31\times10^{22}$ and $2.41\times10^{22}$ cm$^{-2}$, $\mu=1$ and 0.1. This is because of the lowered ionization rate, and reduction of the effective gas-phase formation rate coefficient $R_-$ (Eq.[\ref{eq: Rminusb}]) within the PDR. In the CRZ, the atomic fractions continue to decrease with depth, in accordance with Eq.~(\ref{eq: crbal}) and Fig.~\ref{fig: crz x1x2}, as $\zeta$ decreases and the molecular destruction rate is reduced.

%

The middle panels of Fig.~\ref{fig: fiducial_case_2}
show the abundances of electrons, H$^+$, H$^-$, and H$_3^+$ ions, without (left) and with (right) CR attenuation. As expected, the protons carry the positive charge and the curves for $x_{\rm e}$ and $x_{{\rm H^+}}$ overlap. In the PDR, $x_{\rm e}=2.5\times 10^{-6}$ as given by Eq.(\ref{eq: xe}).  The electron fraction drops in the CRZ, due to the reduced atomic fraction there. The abundance of the H$^-$ intermediary is everywhere proportional to the electron fraction (Eq.[\ref{eq: xHminus}]). Mutual neutralization [R3] and photodetachment [R4] are negligible compared to associative detachment [R2], and $\eta=1$. The H$_3^+$ abundance rises as the H$_2$ density grows. When CR attenuation is included the H$_3^+$ abundance declines sightly with depth as the ionization rate is diminished.


The bottom panels of Fig.~\ref{fig: fiducial_case_2} show the LW photodissociation and CR destruction rates as functions of the total gas column.
For both models,
the onset of molecular self shielding occurs at $N\approx 5\times 10^{19}$~cm$^{-2}$ where $N_{\rm H_2}=2.1\times 10^{15}$,
without CR attenuation, and
$1.5\times 10^{15}$, and $7.2\times 10^{14}$ with CR attenuation (for $\mu=1$ and 0.1), and the photodissociation rate drops with increasing cloud depth. Without CR attenuation
the LW and CR destruction rates are equal at $N=9.3\times 10^{21}$~cm$^{-2}$, near to the H{\small I}-to-H$_2$ transition point. 
The CR destruction rate is slightly larger in the CRZ due to the increase in the $f_d$ branching ratio (Eq.\ref{eq: fd}). With CR attenuation the destruction rates are equal at 
$3.8\times 10^{22}$ and $6.6\times10^{22}$ ~cm$^{-2}$
(for $\mu=1$ and 0.1),
and the CR destruction rates continue to decline within the CRZ.

\subsection{Model Grids}

\begin{table*}
\centering 

\begin{threeparttable}
\caption{Models without cosmic-ray attenuation}
\begin{tabular}{l l l l l l l l}
\hline \hline &&$\log(\zeta_{-16}/n_6)$& -1 & -0.5 & 0 & 0.5 & 1  \\[0.5ex]&$\log(I_{UV}/n_6) $ \\[0.5ex]
\hline
$N_{\rm HI}$ (cm$^{-2}$) &-1 & & 4.1(20) & 1.8(20) & 8.2(19) & 4.8(19) & 2.1(20) \\[0.5ex]
$N_{\rm H_2}$ (cm$^{-2}$) & & & 7.7(19) & 3.5(19) & 1.8(19) & 1.3(19) & 9.2(19) \\[0.5ex]
$N_{\rm tran}$ (cm$^{-2}$) & & & 5.6(20) & 2.5(20) & 1.2(20) & 7.4(19) & 3.9(20) \\[0.5ex]
$f_{\rm{shield}}$ & & & 6.5(-5) & 1.1(-4) & 1.6(-4) & 1.9(-4) & 5.7(-5) \\[0.5ex]
$n_6 \times t_{\rm{eq}}$ (yr)&&& 7.7(7) & 4.3(7) & 2.5(7) & 1.4(7) & 8.0(6) \\[0.5ex]
\\[0.5ex]

&-0.5 && 2.3(21) & 1.1(21) & 5.2(20) & 3.1(20) & 9.5(20) \\[0.5ex]
 &&& 3.6(20) & 2.0(20) & 1.1(20) & 8.5(19) & 4.1(20) \\[0.5ex]
  &&& 3.0(21) & 1.5(21) & 7.4(20) & 4.8(20) & 1.8(21) \\[0.5ex]
   &&& 2.0(-5) & 3.3(-5) & 5.1(-5) & 6.1(-5) & 1.8(-5) \\[0.5ex]
 &&& 7.7(7) & 4.4(7) & 2.5(7) & 1.4(7) & 8.0(6) \\[0.5ex]
  \\[0.5ex]

&0 && 1.0(22) & 5.2(21) & 2.7(21) & 1.6(21) & 3.2(21) \\[0.5ex]
 &&& 1.2(21) & 7.4(20) & 4.8(20) & 3.8(20) & 1.3(21) \\[0.5ex]
  &&& 1.3(22) & 6.7(21) & 3.6(21) & 2.4(21) & 5.8(21) \\[0.5ex]
  &&& 6.4(-6) & 1.0(-5) & 1.6(-5) & 1.9(-5) & 5.7(-6) \\[0.5ex]
 &&& 7.7(7) & 4.4(7) & 2.5(7) & 1.4(7) & 8.0(6) \\[0.5ex]
  \\[0.5ex]

&0.5 && 4.1(22) & 2.1(22) & 1.1(22) & 6.7(21) & 9.0(21) \\[0.5ex]
 &&& 2.9(21) & 2.1(21) & 1.5(21) & 1.2(21) & 3.0(21) \\[0.5ex]
  &&& 4.7(22) & 2.6(22) & 1.4(22) & 9.2(21) & 1.5(22) \\[0.5ex]
   &&& 1.9(-6) & 3.1(-6) & 4.8(-6) & 6.1(-6) & 1.8(-6) \\[0.5ex]
 &&& 7.7(7) & 4.4(7) & 2.5(7) & 1.4(7) & 8.0(6) \\[0.5ex]
  \\[0.5ex]

&1 && 1.5(23) & 8.0(22) & 4.3(22) & 2.5(22) & 2.4(22) \\[0.5ex]
 &&& 6.6(21) & 4.4(21) & 3.2(21) & 3.1(21) & 6.0(21) \\[0.5ex]
  &&& 1.6(23) & 8.9(22) & 5.0(22) & 3.1(22) & 3.6(22) \\[0.5ex]
   &&& 4.4(-7) & 9.7(-7) & 1.6(-6) & 1.7(-6) & 5.4(-7) \\[0.5ex]
 &&& 8.9(7) & 4.4(7) & 2.5(7) & 1.4(7) & 8.0(6) \\[0.5ex]

\hline 

\end{tabular}
\begin{tablenotes}
\item 
The HI, H$_2$, and total 
hydrogen column densities, $N_{\rm tran}$, at the transition points, for each model in the grid shown in Figure~\ref{fig: parameter_grid}. These are models without CR attenuation. The shielding factors and the timescales at the transition points are also listed for each model.
Numbers in parentheses are exponents. For example, $4.1(20)=4.1\times10^{20}$. 
\end{tablenotes}
\label{table: N_tran_no_att}
\end{threeparttable}
\end{table*}

\begin{table*}
\centering 

\begin{threeparttable}
\caption{Models with cosmic-ray attenuation}

\begin{tabular}{l l l l l l l l}

\hline \hline & & $\log(\zeta_{-16}/n_6)$ & -1 & -0.5 & 0 & 0.5 & 1  \\[0.5ex]& $\log(I_{UV}/n_6)$ \\[0.5ex]
\hline
$N_{\rm HI}$ (cm$^{-2}$)& -1 & & 1.4(21) & 4.6(20) & 1.5(20) & 5.7(19) & 4.2(19) \\[0.5ex]
$N_{\rm H_2}$ (cm$^{-2}$) & & & 2.9(20) & 1.1(20) & 3.9(19) & 1.6(19) & 1.4(19) \\[0.5ex]
$N_{\rm tran}$ (cm$^{-2}$) & & & 2.0(21) & 6.8(20) & 2.3(20) & 8.9(19) & 7.0(19) \\[0.5ex]
$f_{\rm {shield}}$ &&& 2.4(-5) & 5.1(-5) & 1.0(-4) & 1.7(-4) & 1.8(-4) \\[0.5ex]
$n_6 \times t_{\rm{eq}}$ (yr) &&& 2.2(8) & 1.0(8) & 4.6(7) & 2.2(7) & 1.2(7) \\[0.5ex]
\\[0.5ex]

&-0.5 && 1.0(22) & 3.9(21) & 1.5(21) & 5.7(20) & 2.8(20) \\[0.5ex]
 &&& 1.4(21) & 6.9(20) & 3.2(20) & 1.5(20) & 8.6(19) \\[0.5ex]
  &&& 1.3(22) & 5.3(21) & 2.1(21) & 8.7(20) & 4.5(20) \\[0.5ex]
   &&& 5.3(-6) & 1.1(-5) & 2.2(-5) & 4.1(-5) & 6.0(-5) \\[0.5ex]
 &&& 3.2(8) & 1.5(8) & 7.0(7) & 3.3(7) & 1.7(7) \\[0.5ex]
  \\[0.5ex]

&0 && 5.5(22) & 2.4(22) & 1.0(22) & 4.3(21) & 1.9(21) \\[0.5ex]
 &&& 3.9(21) & 2.4(21) & 1.4(21) & 8.0(20) & 4.6(20) \\[0.5ex]
  &&& 6.3(22) & 2.9(22) & 1.3(22) & 5.9(21) & 2.9(21) \\[0.5ex]
   &&& 1.2(-6) & 2.5(-6) & 4.9(-6) & 9.6(-6) & 1.6(-5) \\[0.5ex]
 &&& 4.5(8) & 2.1(8) & 1.0(8) & 4.9(7) & 2.4(7) \\[0.5ex]
  \\[0.5ex]

&0.5 && 2.7(23) & 1.2(23) & 5.6(22) & 2.5(22) & 1.1(22) \\[0.5ex]
 &&& 9.9(21) & 6.2(21) & 3.9(21) & 2.6(21) & 1.7(21) \\[0.5ex]
  &&& 2.9(23) & 1.4(23) & 6.4(22) & 3.0(22) & 1.5(22) \\[0.5ex]
   &&& 1.8(-7) & 5.1(-7) & 1.2(-6) & 2.2(-6) & 4.0(-6) \\[0.5ex]
 &&& 8.4(8) & 3.2(8) & 1.4(8) & 6.9(7) & 3.3(7) \\[0.5ex]
  \\[0.5ex]

&1 && 1.2(24) & 5.7(23) & 2.7(23) & 1.2(23) & 5.8(22) \\[0.5ex]
 &&& 1.9(22) & 1.1(22) & 1.0(22) & 6.4(21) & 4.2(21) \\[0.5ex]
  &&& 1.2(24) & 5.9(23) & 2.9(23) & 1.4(23) & 6.6(22) \\[0.5ex]
   &&& 3.2(-8) & 1.4(-7) & 1.7(-7) & 4.8(-7) & 1.0(-6) \\[0.5ex]
 &&& 1.5(9) & 3.8(8) & 2.5(8) & 9.9(7) & 4.4(7) \\[0.5ex]

\hline 

\end{tabular}
\begin{tablenotes}
\item 
The
HI, H$_2$, and total 
hydrogen column densities, $N_{\rm tran}$, at the transition points, for each model in the grid shown in Figure~\ref{fig: parameter_grid_att}. These are models with CR attenuation. The shielding factors and the timescales at the transition points are also listed for each model.
\label{table: N_tran_yes_att}
\end{tablenotes}
\end{threeparttable}

\end{table*}

In Figure~\ref{fig: parameter_grid} we present a $5\times 5$ numerical model grid for the H{\small I} and H$_2$ density profiles, for $I_{\rm LW}/n_6$ ranging from 0.1 to 10 in steps of 0.5 dex (rows), and  $\zeta_{-16}/n_6$ from 0.1 to 10 (columns) also in steps of 0.5 dex. These computations are all for $T=100$~K, without CR attenuation. The middle panel (blue curves) is our fiducial model. In Figure~\ref{fig: parameter_grid} we also plot (dashed curves) the H$_2$ abundances computed using our analytic procedure (\S 3.4). The agreement between the analytic and numerical results is excellent.

%

%

As expected, for fixed $\zeta_{-16}/n_6$, the molecular fractions in the PDRs are reduced as $I_{\rm LW}/n_6$ is increased and photodissociation becomes more effective relative to the fixed formation rates (Eq.~[\ref{eq: x2}]). The transition points (indicated by the red markers) therefore occur at larger gas columns as $I_{\rm LW}/n_6$ is increased. For example, for $\zeta_{-16}/n_6=1$, and for $I_{\rm LW}/n_6$ from 0.1 to 10, the molecular fraction
ranges from $1.26\times 10^{-4}$ to $1.26\times 10^{-6}$ at the cloud edges. The transition columns increase from $1.2\times 10^{20}$ to $5\times 10^{22}$ cm$^{-2}$. The molecular fractions remain unaltered in the CRZs (Eq.[\ref{eq: crbal}] and Figure~\ref{fig: crz x1x2}), where $2x_{\rm H_2}=0.71$ and $x_{\rm HI}=0.29$, independent of $I_{\rm UV}/n_6$.

For increasing $\zeta_{-16}/n_6$ at fixed $I_{\rm LW}/n_6$, the molecular fractions increase in the PDRs but decrease in the CRZs. This is because the H$_2$ formation efficiency via CR ionization is enhanced for larger $\zeta/n$, relative to the fixed photodissociation rates in the PDR (Eq.~[\ref{eq: x2}]). However, in the CRZ the molecular fraction decreases with $\zeta/n$ because removal by CR ionization is more effective than CR driven formation (Eq.~[\ref{eq: crbal}]] and Figure~\ref{fig: crz x1x2}).
The transition columns therefore first {\it decrease} for increasing $\zeta_{-16}/n_6$, but above $(\zeta_{-16}/n_6)_{\rm crit}$=11.84
(see Fig.~\ref{fig: crz x1x2}) a conversion point does not occur and the gas remains predominantly atomic in the CRZ. For example, for $I_{\rm LW}/n_6=1$, and for $\zeta_{-16}/n_6$ from 0.1 to 10, the H$_2$ mass fraction at the PDR edge increases from $8.38\times10^{-6}$ to $8.38\times
10^{-5}$. In the CRZ, it decreases from $0.86$ to $0.52$. For $\zeta_{-16}/n_6$ from 0.1 to 3.16, the transition column decreases from $1.27\times 10^{22}$ to $\sim 2.38\times 10^{21}$. For $\zeta_{-16}/n_6=10$ a transition just barely occurs at 
$5.81\times 10^{21}$~cm$^{-2}$. 

%

In Table \ref{table: N_tran_no_att} we list the H{\small I} and H$_2$ column densities, and the total hydrogen column, at the atomic to molecular transition points ($x_{\rm HI}=2x_{\rm H_2}$) for each of the models in the grid. The required H$_2$ shielding columns range from $\sim 10^{19}$ to a few $10^{21}$~cm$^{-2}$, corresponding to shielding factors, $f_{\rm shield}$, from $1.9\times 10^{-4}$ to $4.4\times 10^{-7}$. We also list the equilibration timescales required to reach the transition points, assuming fully atomic initial states. These range from $n_6\times t_{\rm eq}=8.0\times 10^6$ to $8.9\times 10^7$ yr.

\begin{figure*}
	\centering
	\includegraphics[width=1\textwidth]{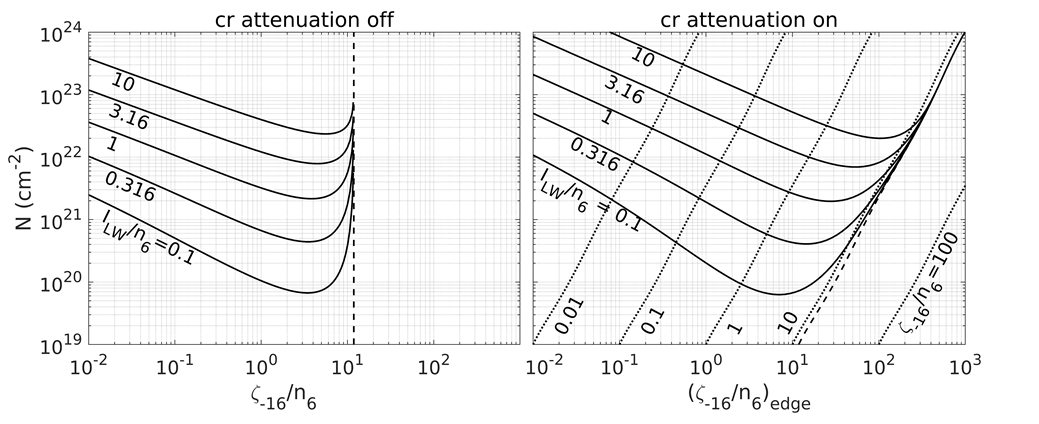} 
	\caption{
Left panel: Total hydrogen gas columns at the atomic to molecular transition points as functions of $\zeta_{-16}/n_6$, for $I_{\rm LW}/n_6$ ranging from 0.1 to 10, for models without CR attenuation. The vertical dashed line indicates the critical ratio ${\zeta_{-16}/n_6}_{\rm crit}=11.84$ (at 100 K). Right panel: Transition columns for models with CR attenuation (for $\mu=1$). The dotted lines are contours of constant local attenuated $\zeta_{-16}/n_6$.  The dashed line is the contour for the critical value. 
		}
		\label{fig: Ntran}
\end{figure*}

In Figure \ref{fig: Ntran} (left panel)
we plot the hydrogen gas columns at which the transition points occur, as functions of $\zeta_{-16}/n_6$ for the various $I_{\rm UV}/n_6$ ratios, for models without CR attenuation. For $I_{\rm UV}/n_6=1$, the transition column decreases from $3.62\times10^{22}$ to a minimum of $2.2\times 10^{21} \rm{cm}^{-2}$, for $\zeta_{-16}/n_6$ from 0.01 to 3.8, and then rises again and diverges at $(\zeta_{-16}/n_6)_{\rm crit}=11.84$
beyond which complete transitions no longer occur. The rise as $\zeta_{-16}/n_6$ approaches the critical value (indicated by the vertical dashed line) is due to the additional contribution of cosmic ray H$_2$ destruction near the inner edge of the PDR, which moderates the shape of the transition profile.

In Figure~\ref{fig: parameter_grid_att}, we display the $5\times 5$ model grid, now including CR attenuation. For these models, we again use Equation~(\ref{eq: Pfit})
assuming Padovani model-$\cal H$ ($q=0.385$) for a normally inclined magnetic field ($\mu=1$). 
The atomic fractions in the PDRs are increased, and the transition points generally occur at greater depths, due to the lowered ionization and reduced H$_2$ formation rates through the PDRs. However, in the CRZs, the molecular fractions are always larger for models with CR attenuation because of the reduced H$_2$ destruction rates, and transitions always occur. 

In Table \ref{table: N_tran_yes_att} we list the H{\small I}, H$_2$, and total hydrogen column densities at the transition points for the model grid shown in Figure~\ref{fig: parameter_grid_att}, as well as the shielding factors and equilibration timescales. As expected, with CR attenuation the required H$_2$ shielding columns are larger, reaching up to $\sim 10^{22}$~cm$^{-2}$, corresponding to $f_{\rm shield}=1.7\times 10^{-7}$.

Figure \ref{fig: Ntran} (right panel)
shows the hydrogen gas columns (solid curves) at the transition points for models with CR attenuation, as functions of the ionization rate to density ratio at the cloud edge, $(\zeta_{-16}/n_6)_{\rm edge}$, for the various $I_{\rm UV}/n_6$
for models including CR attenuation.
For any $I_{\rm UV}/n_6$ the transition columns initially decrease with $(\zeta_{-16}/n_6)_{\rm edge}$ due to the overall enhanced ionization and molecular formation rates in the PDRs. For these branches the atomic to molecular transitions are governed by self-shielding of the H$_2$ against LW photodissociation. 
The (dotted) lines in Figure \ref{fig: Ntran} are contours of constant local attenuated ratios $\zeta_{-16}/n_6$, for our assumed CR attenuation function. The intersections with the solid curves give the local attenuated values of $\zeta_{-16}/n_6$ at the transition points. For example, for $I_{\rm UV}/n_6=1$, and $(\zeta_{-16}/n_6)_{\rm edge}=1.52$, the transition point is at $8.4\times 10^{21}$~cm$^{-2}$ where locally $\zeta_{-16}/n_6=0.1$. The solid curves start rising with $(\zeta_{-16/n_6})_{\rm edge}$ when the transitions occur within the CRZs. The local values of $\zeta_{-16}/n_6$ must then equal $(\zeta_{-16}/n_6)_{\rm crit}=11.84$ (see Figure
\ref{fig: crz x1x2}), so the curves all converge to a line running along this local $\zeta_{-16}/n_6$ contour (close to 10).

\begin{figure}
	\centering
	\includegraphics[width=1\columnwidth]{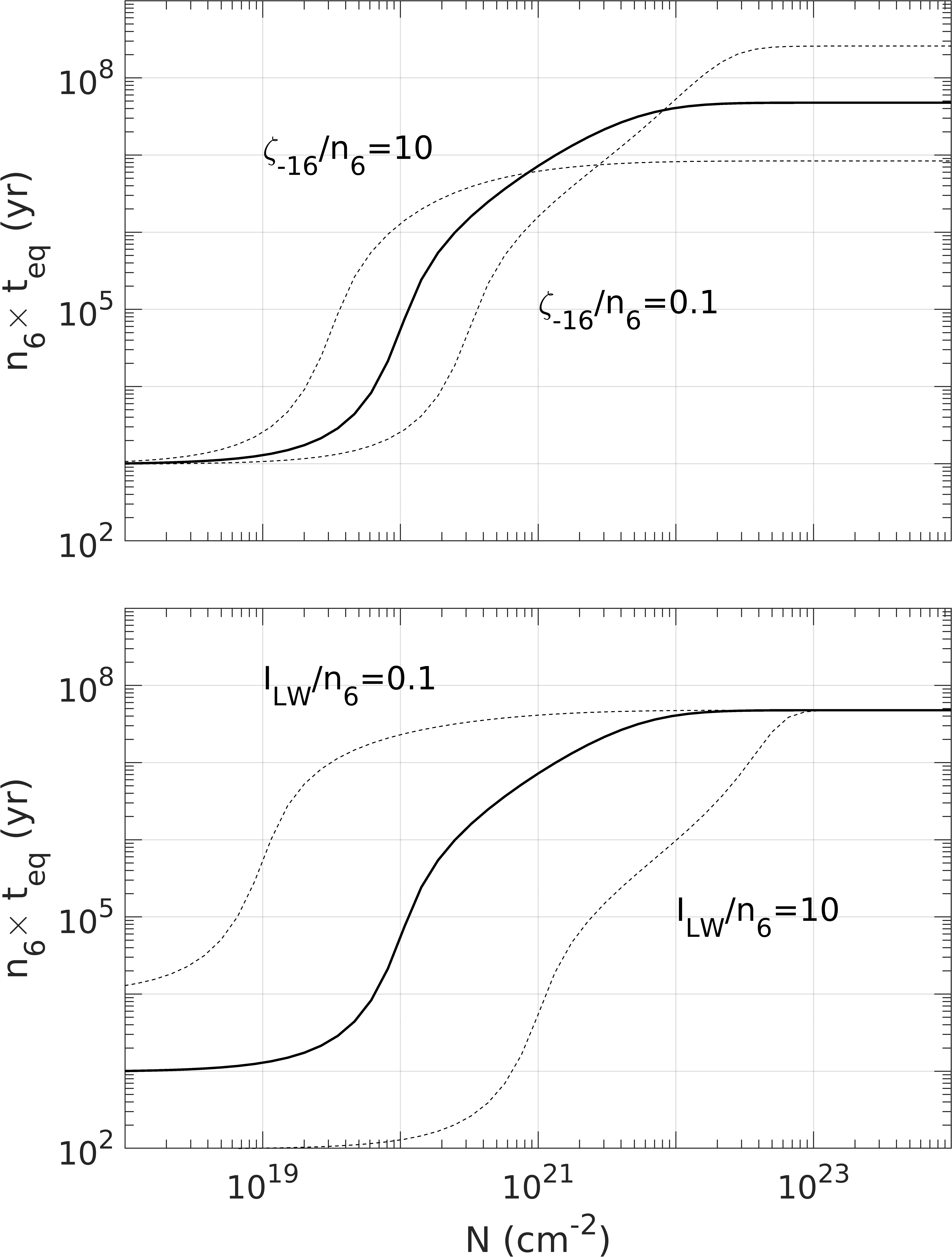} 
	\caption{
Depth-dependent equilibration time-scales for $I_{\rm LW}/n_6=1$ with $\zeta_{-16}/n_6$ from 0.1 to 10 (upper panel), and for  $\zeta_{-16}/n_6=1$, with $I_{\rm LW}/n_6$ from 0.1 to 10 (lower panel).}
    \label{fig: timescale grid}
\end{figure}

\subsection{Depth-dependent time scales}

Figure~\ref{fig: timescale grid} shows the depth-dependent equilibration time-scales, plotted as the product $n_6t_{\rm eq}$ (Eq.~[\ref{eq: teq}]) versus gas column density $N$. These are for models without CR attenuation. We compute $\alpha$ and $\beta$ locally at every depth (rather than using ${\bar \alpha}$ and ${\bar \beta}$). As expected, for fixed $I_{\rm LW}/n_6=1$ (top panel) the time scale approaches the short dissociation limit of $n_6t_{\rm eq}\approx 1.1\times 10^3/I_{\rm LW}$~yr in the PDR at low gas columns, independent of $\zeta_{-16}/n_6$. At large depths in the CRZ, the time scale approaches the long H$_2$ formation limit, with $n_6t_{\rm eq} \approx 4\times 10^7(\zeta_{-16}/n_6)^{-1/2}$~yr. At intermediate gas columns, e.g.~at 10$^{20}$~cm$^{-2}$ in Fig.~\ref{fig: timescale grid}, the time-scale {\it increases} with $\zeta/n$ because elevating the H$_2$ formation rate within the PDR increases the shielding column, thereby reducing the dissociation rate and associated equilibration time. At fixed $\zeta_{-16}/n_6=1$ (bottom panel) the product $n_6t_{\rm eq}$ decreases linearly with $I_{\rm LW}/n_6$ at the cloud edges in the dissociation limit. The time-scale lengthens as self-shielding sets in, finally reaching the long H$_2$ formation time in the CRZ, independent of $I_{\rm LW}/n_6$.

\section{Summary and Discussion}

We have presented an analytic and numerical study of the atomic and  molecular hydrogen density profiles, and (H{\small I}-to-H$_2$) transitions that may occur in dense, dust-free (primoridal) interstellar photodissociation regions (PDRs) and optically thick cosmic-ray zones (CRZs). Our models are highly idealized. We focus on 1D, steady-state, cold, isothermal, constant density gas slabs, irradiated by stellar-type far-UV radiation fields, and fluxes of non-thermal cosmic-ray particles.  Molecular hydrogen formation is by ionization driven gas phase (two-body) chemistry only, primarily via the H$^-$ intermediary, rather than by dust-catalysis as in standard ISM clouds. H$_2$ absorption line self-shielding enables the atomic to molecular transitions, even without any dust opacity. 

For this idealized set-up we solve the rate equations for the depth-dependent abundances of H{\small I} and H$_2$, as well as the trace species, H$^+$, H$_2^+$, H$_3^+$, H$^-$, He, He$^+$, HeH$^+$ and electrons.
The basic parameters in the problem are, (a) the unattenuated LW radiation field intensity $I_{LW}$, (b) the gas density $n$, (c) the cosmic-ray ionization rate $\zeta$, and (d) the gas temperature $T$. For any temperature, the depth dependent abundances and the H{\small I}-to-H$_2$ transition points depend on the two ratios $I_{LW}/n$ and $\zeta/n$. The cosmic-ray driven gas phase H$_2$ formation route may be represented by an effective rate coefficient $R_-$ depending on the ionization fraction (Equations [6]-[9]).

Because gas phase H$_2$ formation is inefficient compared to molecule production on dust-grain surfaces, we focus on relatively dense gas for which the formation rates are enhanced.
Our fiducial model is for $n=10^6$~cm$^{-3}$ at $T=100$~K. 
The illuminating FUV radiation is for a (Pop~III) 10$^5$~K blackbody field normalized to a 
Galactic interstellar LW intensity, $I_{\rm LW}=1$, with LW flux $2.07\times 10^7$ photons~cm~$^{-2}$ s$^{-1}$. The H$_2$ cosmic-ray ionization rate $\zeta=10^{-16}$~s$^{-1}$. We also present results for $I_{LW}/n$ and $\zeta/n$ ranging $\pm 1$ dex around our fiducial case. For this range of parameters, H$^-$ photodetachment is negligible, and the resulting 
H{\small I} and H$_2$ density profiles are insensitive to the IR to UV spectral shapes of the radiation fields. For our fiducial model an H{\small I}-to-H$_2$ transition occurs at a total hydrogen gas column density of $4\times 10^{21}$~cm$^{-2}$, within an equilibration time-scale of $3\times 10^7$ yr. This is comparable to the transition column and H$_2$ formation time for typical dusty Galactic clouds with densities $\sim 10^2$~cm$^{-3}$.

We examine the effects of cosmic-ray energy losses on the H{\small I} and H$_2$ density profiles and transition points. CR attenuation moves the transition points to larger cloud depths due to the reduced ionization fractions and lowered molecular formation efficiencies in the PDRs. However, in the CRZs, the reduced cosmic-ray destruction rates lead to enhanced molecular fractions.  

For our parameter space, the dust-free limit is reached for dust-to-gas ratios $Z^\prime_d\lesssim 10^{-5}$ (relative to Galactic), depending slightly on the ionization rate and gas density. In this limit, gas-phase H$_2$ formation dominates, and dust-gas neutralization processes do not affect the hydrogen-helium chemistry and ionization structures.
Given the probable super-linear dependence of interstellar dust-to-gas mass ratios on the ambient metallicities in differing environments, the dust-free limit may be reached for metallicities $Z^\prime \lesssim 10^{-2}$ (relative to Solar) depending slightly on gas density and ionization rate.

Our theoretical study is relevant for systems such as low metallicity dwarf galaxies at low- and high redshift, possibly dense condensations in the circumgalactic medium of galaxies, and intergalactic filaments, as well as for protoplanetary disks and dust-depleted outflow jets from protostars.  
As for dusty systems, the long H$_2$ formation time scales suggest that cloud evolution, dynamical turbulence, and variability of the radiation sources must be considered for a complete description of the H{\small I} and H$_2$ abundances in specific astrophysical systems.


\vspace{0.2cm}

\noindent
\textbf{Acknowledgements:} We thank Frank Le Petit, David Neufeld, and Marco Padovani for helpful discussions and advice.  
We thank the referee for constructive and helpful comments. This work was supported by the German-Science Foundation via DFG/DIP grant STE/1869-2 GE 625/17-1, by the Center for Computational Astrophysics (CCA) of the Flatiron Institute, and the Mathematics and Physical Sciences (MPS) division of the Simons Foundation, USA.



\section{Appendix A: Metals, Dust, and the Fractional Ionization}

 In Equation~(\ref{eq: xe}) we are assuming that free electrons provided by the ionization of heavy elements are negligible, and that dust grains do not affect the hydrogen-helium chemistry and associated ionization fractions.
 In standard Galactic PDRs free electrons are produced by the photoionization of atomic carbon and other heavy elements with low ionization potentials. For example, in the outer CII zones, photoionization of atomic carbon provides an ionization floor of $x_{\rm e}\sim 2.6\times10^{-4}$ for solar metallicity \citep[e.g.,][]{Sternberg1995}.  Eq.~(\ref{eq: xe}) is then valid for metallicities $Z^\prime \lesssim 0.01 (\zeta_{-16}/n_6)^{1/2}$. Photoionization of heavy elements in dust-free gas could enhance the gas-phase H$_2$ formation efficiency, but we do not consider such effects in this paper. 
 
 Charge transfer between dust grains and ions, or ``dust-assisted recombination", may enhance the neutralization of gas phase ions, including H$^+$ and He$^+$ \citep{Weingartner2001}. The presence of small dust particles (e.g.~polycyclic aromatic hydrocarbons, PAHs) may further alter the gas phase ionization fractions via electron attachment and the formation of negatively charged dust particles \citep{Lepp1988}. For any ion, the dust-assisted recombination rate coefficient depends on the overall dust-to-gas ratio, and the grain charge. We recall that the dust-assisted recombination coefficient is a monotonically decreasing function of the grain charge parameter 
 $\psi= I_{\rm LW}\sqrt{T}/n_e$, where $n_e$ is the electron density. The grain-assisted recombination rate coefficient is maximal for small $\psi$. For Galactic dust abundances the maximal dust-assisted recombination rate for H$^+$ is $\alpha_gn$ where $\alpha_g \sim 10^{-13}$~cm$^3$~s$^{-1}$ \citep{Weingartner2001}. Dust recombination is therefore negligible for $Z^\prime_d \lesssim x_{\rm e}\alpha_{\rm B}/\alpha_g\approx 1.9\times 10^{-4}(\zeta_{-16}/n_6)^{1/2}$ at 100~K.  
 
 For the effects of small grains, we have verified by explicit computation for our model parameter space described in \S 4, that any PAHs are present mainly as neutral or negatively charged particles. We calculated the relative PAH$^+$/PAH/PAH$^-$ abundances using the rate coefficients for recombination and attachment with electrons, and charge transfer reactions with H and H$^+$, as summarized in \cite{Wolfire2003} (see their Appendix C2), and the photoionization and photodetachment cross sections recommended by \cite{Lepp1988}. For a characteristic PAH abundance $x_{\rm PAH,\odot}=6\times 10^{-7}$ for $Z^\prime_d=1$ \citep{Lepp1988,Weingartner2001} we find that the PAHs do not alter the hydrogen-helium chemistry and ionization structure for $Z^\prime_d\lesssim 10^{-5}$. This can also be seen analytically. The dominant neutralization process is attachment (${\rm PAH + e \rightarrow PAH^- + \nu}$), with a rate coefficient $\kappa_- = 1.34\times 10^{-6}$~cm$^3$~s$^{-1}$. This reaction is negligible compared to radiative recombination with protons for $Z^\prime_d \lesssim x_{\rm e}\alpha_{\rm{B}}/x_{\rm{PAH},\odot}\kappa_- = 2.4\times 10^{-5}(\zeta_{-16}/n_6)^{1/2}$. We conclude that for $\zeta_{-16}=n_6=1$ dust-grain neutralization and H$_2$ formation on dust surfaces (as discussed in \S 3.1) are, coincidentally, both negligible at dust-to-gas ratios $Z^\prime_d\sim 10^{-5}$.

\section{Appendix B: Comparison to Sternberg et al.~2014}

As discussed in \citetalias{Sternberg2014}, in dusty {\it optically thick} PDRs  the total H{\rm I} column density maintained by photodissociation for beamed radiation is 
\begin{equation}
\label{eq: apN1tot}
    N_{_{\rm HI,\rm tot}}^{\rm LW} = \frac{1}{\sigma_g} {\rm ln}[\frac{\alpha G}{2} + 1] = \frac{1}{\sigma_g}{\rm ln}\bigl[\frac{1}{2}\frac{\sigma_g {\bar F}_{\nu,{\rm LW}}W_{g,{\rm tot}}}{R_{\rm dust}n} + 1\bigr] \ \ \ ,
\end{equation}
where the dimensionless parameters
\begin{equation}
    \alpha \equiv \frac{D_0}{R_{\rm dust}n} \ \ \ ,
\end{equation}
and
\begin{equation}
    G\equiv \frac{\sigma_g}{\sigma_d}W_{g,{\rm tot}} \ \ \ .
\end{equation}
In these expressions, ${\bar F}_{\nu,{\rm LW}}$ is the mean LW band flux-density (photons cm$^{-2}$ s$^{-1}$ Hz$^{-1}$), $\sigma_d$ is the total H$_2$ photodissociation cross section (cm$^2$~Hz), $D_0=\sigma_d{\bar F}_{\nu,{\rm LW}}$ is the free-space photodissociation rate (s$^{-1}$), $W_{g,{\rm tot}}$ (Hz) is the total H$_2$-dust limited dissociation bandwidth (see \citetalias{Sternberg2014}), $\sigma_g$ is the LW dust continuum absorption cross section (cm$^2$), $R_{\rm dust}$ is the dust-grain H$_2$ formation rate coefficient (cm$^3$~s$^{-1}$) and $n$ is the gas density (cm$^{-3}$). The dimensionless parameter, $\alpha$ is the ratio of the free-space dissociation rate to the H$_2$ formation rate, and $G$ (also dimensionless) is the dust-opacity averaged H$_2$ self-shielding factor (see Eq.~45 in \citetalias{Sternberg2014}).   

In the strong field limit, $\alpha G \gtrsim 1$, and
\begin{equation}
    N_{_{\rm HI,\rm tot}}^{\rm LW} \approx \frac{1}{\sigma_g}{\rm ln}\bigl[\frac{1}{2}\frac{\sigma_g {\bar F}_{\nu,{\rm LW}}W_{g,{\rm tot}}}{R_{\rm dust}n}\bigr] \ \ \ .
\end{equation}
In this limit, the H{\small I}-dust opacity $\sigma_gN_{\rm HI,{\rm tot}}^{\rm LW}$ associated with the photodissociated H{\small I} column is large. The H{\small I} column is only logarithmically sensitive to $\alpha G$ because the H{\small I} column is self-limited by the H{\small I}-dust absorption that competes with H$_2$ photodissociations. The dominating H{\small I}-dust opacity leads to sharp H{\small I}-to-H$_2$ transitions due to the exponential dust reductions of the LW flux (see Fig.~7 in \citetalias{Sternberg2014}).

In the weak-field limit, $\alpha G \ll 1$, and
\begin{equation}
\label{eq: weakF}
     N_{_{\rm HI,\rm tot}}^{\rm LW} \approx \frac{1}{2}\frac{ {\bar F}_{\nu,{\rm LW}}W_{g,{\rm tot}}}{R_{\rm dust}n} \ \ \ .
\end{equation}
In this limit H{\small I}-dust opacity is negligible, and the H{\small I} column is linear with $\alpha G$, i.e.~with the incident LW flux. 
The H{\small I}-to-H$_2$ transitions are controlled by H$_2$ self-shielding in non-overlapping absorption lines and the transitions are gradual (not sharp). In the weak-field limit a large fraction of the photodissociated H{\small I} column is built up past the transition points inside the molecular zones (see Fig.~7 in \citetalias{Sternberg2014}).

We now consider dust-free systems. For these, $\sigma_g \rightarrow 0$, and $W_{g,{\rm tot}} \rightarrow W_{d,{\rm tot}}$, where $W_{d,{\rm tot}}$ is the dust-free H$_2$ dissociation bandwidth. Thus, $G\rightarrow 0$.
Furthermore, $R_{\rm dust}$ is replaced with a non-vansihing $R_-$. So for dust-free conditions $\alpha G$ is small by definition, and
\begin{equation}
     N_{\rm HI,{\rm tot}}^{\rm LW} \approx \frac{1}{2}\frac{ {\bar F}_{\nu,{\rm LW}}W_{d,{\rm tot}}}{R_-n} \ \ \ .
\end{equation}
This is formally analogous to the weak-field limit (eq.~[\ref{eq: weakF}]) for dusty clouds for which (a) H{\small I} dust opacity is negligible (b) the transitions are controlled by H$_2$ self-shielding, and (c) the H{\small I} is linearly proportional to the incident LW flux.  However, for dust-free clouds a different distinction between weak and strong fields remains. For $N_{_{\rm HI,\rm tot}}^{\rm LW} \gtrsim 10^{22}$~cm$^{-2}$, which is the H$_2$ column density required for complete line-overlap, the H{\small I}-to-H$_2$ transitions are sharp due to the exponential cutoff of the dissociation rates. This is the strong-field limit for dust-free clouds. For $N_{_{\rm HI,\rm tot}}^{\rm LW} \lesssim 10^{22}$~cm$^{-2}$, the transitions occur before the absorption lines overlap. The conversions are then gradual with significant H{\small I} in the molecular zones up to the point where the lines finally do overlap and the LW band photons are fully absorbed. This is the weak-field limit for dust-free clouds. 
As discussed in \S 3.1 and Appendix A, for $\zeta_{-16}/n_6\sim 1$ the dust-free limit is reached for dust-to-gas ratios $Z^\prime_d \lesssim 10^{-5}$, and overall metallicities $Z^\prime \lesssim 10^{-2}$, relative to characteristic Galactic values.

\end{document}